\documentclass[letterpaper, appendixfloats]{emulateapj}

\pdfoutput=1

\usepackage{epsfig}
\usepackage{lscape}
\usepackage{natbib}
\usepackage{array}
\usepackage{appendix}
\def\plotone#1{\centering \leavevmode
\includegraphics[clip=, width=.85\columnwidth]{#1}}

\newcommand{\cN}[1]{\mathcal{N}}

\def\gsim{\;\rlap{\lower 2.5pt
 \hbox{$\sim$}}\raise 1.5pt\hbox{$>$}\;}
\def\lsim{\;\rlap{\lower 2.5pt
   \hbox{$\sim$}}\raise 1.5pt\hbox{$<$}\;}

\def\lsun{{L_\odot}}

\begin{document}


\title{%
Habitable Climates: The influence of eccentricity\\
}

\author{
Courtney D. Dressing\altaffilmark{1},
David S. Spiegel\altaffilmark{1,2},
Caleb A. Scharf\altaffilmark{3,4},
Kristen Menou\altaffilmark{2,4},
Sean N. Raymond\altaffilmark{5,6},
}
 
 \affil{$^1$Department of Astrophysical Sciences, Princeton
 University, Peyton Hall, Princeton, NJ 08544}
\affil{$^2$Kavli Institute for Theoretical Physics, UCSB, Santa Barbara, CA 93106-4030}
 \affil{$^3$Columbia Astrobiology Center, Columbia Astrophysics
 Laboratory, Columbia University, 550 West 120th Street, New York, NY
 10027}
 \affil{$^4$Department of Astronomy, Columbia University, 550 West
 120th Street, New York, NY 10027}
\affil{$^5$Universit\'{e} Bordeaux, Observatoire Aquitain des Sciences
 de l'Univers, 2 rue de l'Observatoire, BP 89, F-33271 Floirac Cedex, France}
\affil{$^6$CNRS, UMR 5804, Laboratoire d'Astrophysique de Bordeaux, 
2 rue de l'Observatoire, BP 89, F-33271 Floirac Cedex, France}
 \vspace{0.5\baselineskip}
 
 \email{courtney@astro.princeton.edu, dsp@astro.princeton.edu,
   caleb@astro.columbia.edu,\\ kristen@astro.columbia.edu,
   raymond@obs.u-bordeaux1.fr}

\begin{abstract}
In the outer regions of the habitable zone, the risk of transitioning
into a globally frozen ``snowball'' state poses a threat to the
habitability of planets with the capacity to host water-based
life. Here, we use a one-dimensional energy balance climate model
(EBM) to examine how obliquity, spin rate, orbital eccentricity, and
the fraction of the surface covered by ocean might influence the onset
of such a snowball state.  For an exoplanet, these parameters may be
strikingly different from the values observed for Earth.  Since, for
constant semimajor axis, the annual mean stellar irradiation scales
with $(1-e^2)^{-1/2}$, one might expect the greatest habitable
semimajor axis (for fixed atmospheric composition) to scale as
$(1-e^2)^{-1/4}$. We find that this standard simple ansatz provides a
reasonable lower bound on the outer boundary of the habitable zone,
but the influence of both obliquity and ocean fraction can be profound
in the context of planets on eccentric orbits. For planets with
eccentricity 0.5, for instance, our EBM suggests that the greatest
habitable semimajor axis can vary by more than 0.8 AU (78\%!)
depending on obliquity, with higher obliquity worlds generally more
stable against snowball transitions.  One might also expect that the
long winter at an eccentric planet's apoastron would render it more
susceptible to global freezing. Our models suggest that this is not a
significant risk for Earth-like planets around Sun-like stars, as
considered here, since such planets are buffered by the thermal
inertia provided by oceans covering at least 10\% of their surface.
Since planets on eccentric orbits spend much of their year
particularly far from the star, such worlds might turn out to be
especially good targets for direct observations with missions such as
TPF-Darwin.  Nevertheless, the extreme temperature variations achieved
on highly eccentric exo-Earths raise questions about the adaptability
of life to marginally or transiently habitable conditions.
\end{abstract}


\section{Introduction}
\label{sec:intro}
There are now more than 460 extrasolar planets known.\footnote{See
http://exoplanet.eu, http://exoplanets.org} Selection effects favor
the discovery of massive giant planets orbiting very close to their
parent stars, but advances in imaging and spectroscopic capabilities,
longer baselines for observation, and missions such as \emph{CoRoT}
and \emph{Kepler} should accelerate the rate of discovery of less
massive planets in longer period orbits in the near future.  In
particular, microlensing observations have already detected planets
less than ten times as massive as the Earth at distances as great as
2.6~AU \citep{ beaulieu_et_al2006, bennett_et_al2008}, and have
discovered a potential analog to our solar system
\citep{gaudi_et_al2008b} that might allow a habitable planet on a
stable orbit \citep{malhotra+minton2008}. Both \emph{CoRoT} and
\emph{Kepler} promise to further increase the number of detected
terrestrial exoplanets \citep{baglin2003,borucki_et_al2003,
borucki_et_al2007, borucki2010}.  The observed secondary eclipse of HAT-P-7b by
\emph{Kepler} demonstrates that it should be capable of detecting
transits of Earth-size planets \citep{borucki_et_al2009}.
Additionally, the \emph{CoRoT} team recently announced the detection
of the $\sim$1.7$R_\earth$ exoplanet COROT-Exo-7b orbiting a K0 star
in the constellation Monoceros \citep{rouan_et_al2009,
leger_et_al2009, bouchy_et_al2009,fressin_et_al2009} and the MEarth
Project detected the transit of the 6.55~M$_\earth$ planet GJ 1214b
\citep{charbonneau_et_al2009}.  Although these planets are in
extremely close orbits ($a = 0.017$~AU, P = 0.85 days for COROT-Exo-7b;
$a = 0.014$~AU, P = 1.58 days for GJ 1214b) and are unlikely to be
habitable, their discoveries represent a tremendous advance in planet
detection capability.  As the \emph{CoRoT}, \emph{Kepler}, and MEarth
projects continue, even less massive terrestrial planets will surely
be discovered in orbits with greater semimajor axes.  Once these
potentially habitable terrestrial planets are discovered, researchers
will be able to determine the radii, orbital semimajor axes, and
masses of the planets, but current techniques are insufficient to
constrain their obliquities and rotation rates
\citep{valencia_et_al2006, valencia_et_al2007, adams_et_al2008}.
Although transit measurements might place constraints on the
eccentricities of some transiting planets \citep{barnes2007,
ford_et_al2008}, radial velocity measurements of exact Earth analogs
will be extremely challenging, and so the eccentricities of most such
planets will remain undetermined in the near future.  This paper
attempts to quantify the effects of orbital parameters such as
eccentricity on planetary habitability in order to prepare us to draw
inferences about the habitability of yet-to-be-detected terrestrial
planets even if their eccentricities are not well constrained.

However, current surveys of extrasolar planets indicate that the
near-zero eccentricities seen in the Solar System are not
necessarily typical and that many planets have significantly higher
eccentricities \citep{udry+santos2007}.  Of the exoplanets with
measured eccentricities, $\sim$40\% are on more eccentric orbits than
Pluto (${e = 0.2488}$) and $\sim$10\% are on orbits with
eccentricities $>$0.5.  This suggests that current views of
habitability that focus on direct Earth analogs in near-circular
orbits might consider only a small subset of potentially habitable
worlds.  It therefore seems prudent to expand our study of
habitability to encompass a wide range of orbital parameters.

The study of planetary habitability began decades prior to the
detection of exoplanets with the classic work of \citet{dole1964} and
\citet{hart1979}.  In the last two decades, this topic has been
revisited with increasing frequency, beginning with the work of
\citet{kasting_et_al1993}, who found conservative limits for the
Earth's liquid water habitable zone between 0.95~AU and 1.37~AU.  In
recent years, various studies have applied the tools and techniques
used to study the Earth's climate to simulations of planets around
other stars.  Investigations by \citet{williams+pollard2003},
\citet{williams+kasting1997} and \citet{spiegel_et_al2009} have shown
that, while variations in the polar obliquity angle can alter the
distance from the star at which a planet becomes too cold to be
habitable, planets with high obliquities are not necessarily less
habitable than planets with low obliquities.
\citet{williams+pollard2002} also considered the effect of
eccentricity on habitability but only in the context of Earth twins at
1~AU.

\citet{spiegel_et_al2008, spiegel_et_al2009} have
analyzed the effect of changes in semimajor axis, rotation rate,
obliquity, and ocean coverage on the temperature of a generic
terrestrial planet, but not variations in
eccentricity.  This paper attempts to fill in the void between the
work of \citet{spiegel_et_al2009} and \citet{williams+pollard2002} by
varying the eccentricity of model planets that are more diverse than
the Earth-twin used by \citet{williams+pollard2002}.  In addition, we
also conduct a sensitivity study to determine which parameters have
the strongest effect on planetary temperatures, so as to 
quantify the degree to which uncertainty in parameter measurement
translates to uncertainty in climate.

Although some planets may be habitable at all latitudes during their
entire orbit, other planets, like the Earth, might be only partially
habitable.  These planets could therefore transition to a ``snowball
state'' if small changes in insolation or atmospheric composition
cause part or all of the planet to freeze. During a ``snowball
transition,'' the formation of snow or ice increases the albedo of the
planet and the planet consequently becomes even colder. If the
positive feedback loop between ice formation and increased albedo
continues, the entire planet may become frozen and trapped in a
snowball state.  As discussed in Section~\ref{ssec:OHZ}, the
accumulation of greenhouse gases in the atmosphere while the planet is
frozen may warm the surface sufficiently to allow the planet to
eventually exit the snowball state.  The transition from a snowball
state to a partially habitable state is beyond the scope of this
paper, but an investigation is pursued in \citet{pierrehumbert2005} as
well as in a companion paper \citep{spiegel_et_al2010b}.

In this paper, based on a simple energy balance model (EBM) treatment,
we determine that obliquity, eccentricity, and ocean fraction can
together have a very strong influence on the orbital location of the
snowball transition.  For instance, for models with an Earth-like
atmosphere and eccentricity 0.5 (which might not be extreme by
extrasolar standards), we find that the maximum habitable semimajor
axis can extend to 1.90~AU or be as close to the star as 1.07~AU,
depending on obliquity and ocean fraction.  As a result, for $e = 0.5$,
the standard ansatz for the outer boundary of the habitable zone
derived from considering only the annual mean flux
\citep{barnes_et_al2008} can be off by more than 78\%.  Altering the
azimuthal obliquity (the degree of alignment of periastron with the
solstices) does not have a significant effect on where the snowball
transition occurs for low eccentricity planets, but the transition for
high eccentric planets is pushed out significantly for azimuthal
obliquity $\sim$30$^\circ$.  More generally, planets in higher
eccentricity orbits display more latitudinal variation and seasonal
variation in habitability than planets in near-circular orbits.
Because all of the models in this study incorporate an Earth-like
atmosphere, simultaneous variations of atmospheric composition or
other factors in combination with the parameters considered in this
study may produce a more complicated picture of general planetary
habitability.

In Section \ref{sec:milcyc} we discuss several important factors that
influence the Earth's climate on long time-scales. We explain the
setup of our model in Section \ref{sec:modelset} and discuss the
validation of the model in Section \ref{sec:modelval}.  In Section
\ref{sec:results} we present our results.  We then conclude in Section
\ref{sec:conc} and consider the implications of our findings on
planetary habitability.

\section{Generalized Milankovitch Cycles}
\label{sec:milcyc}
\citet{milankovitch1941} realized that the long-term climate behavior
of the Earth could in part be explained by considering the combined
effects of obliquity, orbital eccentricity, and precession.  Each of
these orbital elements changes on multiple, nonconstant timescales
known as Milankovitch cycles and the combination of these variations
alters the climate of the Earth by increasing or decreasing the solar
insolation received by the Earth at a given latitude.  For example,
increasing the Earth's obliquity increases the annually averaged
insolation received by the poles and decreases the annually averaged
insolation received by the equator; both effects act to decrease the
latitudinal temperature gradient." However, the Earth's obliquity is
largely stabilized by the Moon and varies by only ${\pm1.2^\circ}$ on
timescales of $\sim$41~kyr \citep{laskar_et_al1993, berger1976,
berger1978}.  Much higher variations in obliquity are expected for
planets without large moons: the obliquity of Mars varies between
${14.9^\circ}$ and ${35.5^\circ}$ \citep{ward1974} and numerical
models suggest that the Earth would experience obliquity oscillations
between ${0^\circ}$ and ${85^\circ}$ in the absence of the Moon
\citep{laskar_et_al1993, laskar+robutel1993}.  Even in the absence of
large moons, however, the obliquity of a quickly rotating planet
($\lesssim$8-hour day) would probably be self-stabilized by the fast
rotation rate of the planet \citep{ward1982, laskar_et_al1993}.

In addition to obliquity, there is also a Milankovitch cycle governing
precession on shorter timescales of $\sim$19~kyr and $\sim$23~kyr
\citep{berger1976, berger1978}.  Over time, the slow shift of the
direction of the Earth's rotation axis due to precession of both the
spin axis and the orbital ellipse alters the position of solstices and
equinoxes with respect to apoastron and periastron.  In this paper, we
consider this effect by varying the azimuthal obliquity angle
$\theta_a$ of our model planets (defined in Section \ref{sec:modelset}
as the angle between the position of the planet at periastron and the
position of the planet at the northern winter solstice).  Due to the
ice albedo effect, the hemisphere that is tilted toward the star at
apoastron---which has a shorter winter (i.e., a shorter period of high
albedo)---absorbs more integrated stellar energy per year than does
the hemisphere that is tilted away at apoastron.  Accordingly, the
greatest temperature asymmetry between the northern and southern
hemispheres is produced when periastron is aligned with a solstice
(${\theta_a = 0^\circ}$ or ${\theta_a = 180^\circ}$).  Conversely,
both hemispheres absorb equal annually averaged stellar irradiation
when periastron is aligned with an equinox (${\theta_a = 90^\circ}$ or
${\theta_a = 270^\circ}$).

Finally, there is a Milankovitch cycle for eccentricity.  The Earth's
orbital eccentricity is currently nearly circular (${e = 0.0167}$),
but varies slowly up to $\sim$0.06 over long timescales of $\sim$100
and $\sim$400~kyr \citep{berger1976, berger1978}.  As shown in Figure
\ref{fig:rel}, the annual mean flux~${\left< F \right>}$ scales
as~${(1-e^2)^{-1/2}}$ so that flux increases with increasing
eccentricity for a given semimajor axis. Increasing the eccentricity,
therefore, accentuates the ratio of the irradiating flux at periastron
to that at apoastron (${F_{\rm max}/F_{\rm min} \propto
((1+e)/(1-e))^2}$), and slightly increases the annually averaged
irradiation.  The ratio ${F_{\rm max}/F_{\rm min}}$ can be quite
substantial for highly eccentric orbits, exceeding $10^2$ for
$e>0.82$, which could, depending on a planet's thermal inertia and
redistribution of energy, cause dramatic seasonal temperature swings.
We present climate models of planets with high orbital eccentricity in
Section \ref{ssec:OHZ}, and refer readers to a companion paper
\citep{spiegel_et_al2010b} for a discussion of scenarios in which the
eccentricity of an Earth-like planet might be excited to large values.

Over time, the combined changes in obliquity, precession, and
eccentricity, together with nonlinear amplifications, can dramatically
alter an exo-Earth's climate, leading alternately to periods of
glaciation and of deglaciation as has been shown in the case of the
Earth \citep{berger_et_al2005, crucifix_et_al2006,laskar_et_al1993,
loutre_et_al2004, quinn_et_al1991}.

\section{Model Setup}
\label{sec:modelset}
In this study, we investigate the temperature of a planet using the
same one-dimensional time-dependent energy balance model introduced in
\citet{spiegel_et_al2008} and further explored in
\citet{spiegel_et_al2009}. The model, which is similar to the more
Earth-centric model used by \citet{suarez+held1979}, treats the
meridional transport of heat as diffusion driven by the zonal mean
temperature gradient:
\begin{equation}
C \frac {\partial T[x,t]}{\partial t} - \frac{\partial}{\partial x} \left( D(1-x^2)\frac{\partial T[x,t]}{\partial x}\right)+I = S(1-A) \, .
\label{eq:surftemp}
\end{equation}
This equation describes the evolution of the temperature $T$ at
location $x \equiv \sin \lambda$, where $\lambda$ is the latitude, as
a function of an effective heat capacity $C$, a diffusion coefficient
$D$, and an albedo $A$. The net radiative energy flux in a latitude
band is determined by the relationship between the energy received due
to the diurnally averaged stellar flux $S$ and the energy lost due to
infrared emission $I$.  One-dimensional EBMs such as this model
provide a reasonable approximation of seasonal mean temperatures for
planets that rotate sufficiently quickly relative to their orbital
frequency \citep{showman_et_al2009}.  In all of our models we assume
that the planet orbits a Sun-like star, so the stellar flux $S$ is
equivalent to that from a 1~$M_\sun$, 1~$L_\sun$ star.  The effective
heat capacities of the atmosphere over land ($C_l$), over the
wind-mixed surface layer of the ocean ($C_o$), and over ice ($C_i$)
are the same as in \citet{spiegel_et_al2008,spiegel_et_al2009} and
\citet{williams+kasting1997} and are shown in Table \ref{tab:heatcap}.

Previous work \citep{spiegel_et_al2008} explored three sets of
infrared cooling radiation functions and albedo functions.  Of the
three models tested in that paper, Model 2 produced climates most
similar to those on current Earth and is the one used in the current
study:
\begin{equation}
I[T]=\frac{\sigma T^4}{1+0.5925 \left( \frac{T}{273 \mathrm{\ K}}\right)^3} \, \,  ,
\label{eq:ircool}
\end{equation}
where $\sigma$ is the Stefan-Boltzmann constant. The denominator of
equation \ref{eq:ircool} would instead be 1 if the atmosphere had no
opacity to outgoing infrared radiation; the functional form used for
this denominator represents the strength of the greenhouse effect, and
is a reasonable approximation of the greenhouse for present-Earth
conditions \citep{spiegel_et_al2008}.  In order to account for the
higher reflectivity of ice and snow while using a simple functional
form, we take the albedo to be constant and low ($\sim$0.28) at high
temperatures, constant and high ($\sim$0.77) at low temperatures, and
to vary smoothly in between:
\begin{equation}
A[T]=0.525-0.245\tanh \left( \frac{T-268 \mathrm{\ K}}{5 \mathrm{\ K}} \right) \, .
\label{eq:albedo}
\end{equation} 
The smooth hyperbolic tangent formulation is chosen to handle the
phase transition from water to ice at 273~K in order to avoid the
small ice-cap instabilities seen in models with a discontinuity in the
albedo function at 273 K \citep{held_et_al1981}.

As explained in \citet{spiegel_et_al2008}, the fiducial diffusion
coefficient follows the form of \citet{williams+kasting1997}
and is taken to be ${D_{\rm fid} = 5.394 \times 10^2
{\rm~erg~cm^{-2}~s^{-1}~K^{-1}} \times
(\Omega_p/\Omega_{\earth})^{-2}}$, where $\Omega_p$ and $\Omega_\earth$
are the rotation rates of the exoplanet and the Earth, respectively.
Its value increases with decreasing planetary angular spin frequency.
We note that while the ``thermal Rossby Number'' scaling argument of
\citet{farrell1990} supports this dependence of $D$ on $\Omega_p$,
\citet{delgenio_et_al1993} and \citet{delgenio+zhou1996} find that the
effective diffusivity of slowly rotating planets might not follow such
a simple scaling relationship. The reader is directed to
the recent review by  \citet{showman_et_al2009} for an 
in-depth discussion of the relationship
between planetary rotation rate and effective diffusivity.

The model is solved by relaxation on a grid of 145 uniformly spaced
latitude points using a time-implicit numerical scheme and an adaptive
time-step, as described in \citet{spiegel_et_al2008} and
\citet{hameury_et_al1998}.  We typically initialize the planet at the
northern winter solstice, but for some runs we vary the azimuthal
obliquity $\theta_a$ to change the initial season.  The initial
temperature in all cases is uniform across the surface of the planet
and is chosen to be at~least~350~K to minimize the likelihood that
models evolve into ice-covered snowball Earths.  Similar initial
conditions were also used by \citet{kasting_et_al1993} and
\citet{spiegel_et_al2008, spiegel_et_al2009} to avoid ``cold start''
planets.  As explained in the Appendix of \citet{spiegel_et_al2008},
as long as the initial planet temperature is significantly warmer than
273~K, our model relaxation studies indicate that within 130 years of
model evolution (and sometimes far less), the final state of the
planet is independent of the initial conditions.  If the initial
temperature is ${T\lesssim273 \rm~K}$, however, the planet can quickly
transition to a snowball state due to water-ice albedo feedback.  Our
choice of initial conditions should therefore lead to more optimistic
results for the location of the snowball transition.

The purpose of this study is to examine the influence of various
orbital and planetary parameters on planetary habitability and to
determine the sensitivity of a planet's habitability to changes in
those parameters. Figure \ref{fig:orbparam} portrays a schematic
diagram indicating the relevant angles; here, we give a detailed
description of parameters:
\begin{enumerate}
 \item{\emph{Eccentricity $e$.} We vary the eccentricity of our model
   planets from ${e = 0}$ to ${e = 0.90}$.}
 \item{\emph{Polar Obliquity Angle $\theta_p$.} This is the angle
   between the rotation axis of the planet and the normal to the plane
   of rotation.  Because our model planets are symmetric, we restrict
   this angle to between ${\theta_p = 0^\circ}$ and ${\theta_p =
   90^\circ}$.  Values between ${\theta_p = 90^\circ}$ and ${\theta_p =
   180^\circ}$ simply reverse the designation of the identical northern
   and southern hemispheres. The wide range of polar obliquities is 
   appropriate given the variety of polar obliquities in our own solar
   system and the range of obliquities predicted by simulations such
   as \citet{agnor_et_al1999, laskar_et_al1993, laskar+robutel1993}.
   In particular, \citet{kokubo+ida2007} suggest that polar obliquities
   near 90$\degr$ may actually be more common than polar obliquities
   near 0$\degr$.}
   
 \item{\emph{Azimuthal Obliquity Angle $\theta_a$.} This is the angle
   between the projection of the rotation axis of the planet onto the
   plane of rotation and the line between the star and the periastron
   position of the planet.  Variations in this angle change the
   initial season.  The model is always initialized at periastron, and
   periastron coincides with northern winter for most models because
   most models have ${\theta_a = 0}$. Once the model reaches a
   periodic state, the average temperature of the planet will be
   greater at periastron than at apoastron.  Consequently,
   initializing the models at periastron is a relatively conservative
   choice because the average planetary temperature will decrease as
   the planet approaches apoastron and the planet could enter a
   snowball state more quickly than a planet that was initialized with
   an average global surface temperature of $\gtrsim 350$~K at
   apoastron.}
 \item{\emph{Rotation Rate $\Omega_p$.} Because our parametrization of
   the diffusion coefficient depends inversely on the square of the
   rotation rate, increasing the rotation rate is equivalent to
   reducing the efficiency of latitudinal heat transport.  In this
   study we consider planets with 8-hour days (${\Omega_p = 3
   \Omega_\earth}$, ${D = (1/9)D _{\rm fid}}$), 24-hour days (${\Omega_p =
   \Omega_\earth}$, ${D = D_{\rm fid}}$) and 72-hour days (${\Omega_p =
   (1/3)\Omega_\earth}$, ${D = 9D_{\rm fid}}$).}
 \item{\emph{Semimajor Axis a.} We begin each set of models by using a
   simple scaling approximation as  an ansatz about the location
   of the outer boundary of climatic habitability and then run a
   series of models with semimajor axes near that value to locate and
   refine the position at which a model planet first becomes fully
   ice-covered year-round.}
 \item{\emph{Ocean Fraction $f_o$.} This parameter determines the
   ratio of land to ocean found on the surface of the planet.  Since
   the atmosphere over the wind-mixed layer of the ocean has a much
   higher effective thermal inertia than the atmosphere over land,
   waterworld planets with high ocean fractions experience less
   dramatic temperature variations than do desert worlds with low
   ocean fractions.  Although the wind-mixed layer of the ocean varies
   in depth from a few meters to a few hundred meters or more
   \citep{hartmann1994}, we assume a depth of 50~m for the
   study. Increasing the depth of the wind-mixed layer would enhance
   the effective surface heat capacity and lengthen the timescale on
   which temperature changes occur.  Theoretical simulations
   by \citet{marotzke+botzet2007} indicate that the depth of the
   wind-mixed layer increases dramatically as the Earth freezes over,
   so our use of a constant 50-m wind-mixed layer may increase the
   likelihood that a planet will transition to a snowball state.
   Indeed, our most Earth-like model ($a = 1$~AU, $e = 0$, $\theta_p =
   23.5\degr$, $\theta_a = 0\degr$, $f_o = 70\%$) transitions to a
   snowball state when the stellar luminosity has been reduced to
   0.995$\lsun$; at 0.996$\lsun$, more than 29\% of the surface is
   covered by ice. This is comparable to the maximum stable ice cover
   of 30\% reported by \citet{north1975}, but significantly below the
   value of 55\% found by \citet{voigt+marotzke2009}. In this
   study, we present results from simulations with ocean coverage
   ranging between 10\% and 90\%, but most of our models incorporate
   an Earth-like 70\% ocean fraction (${f_o = 0.7}$).  Considering a
   range of ocean fractions is important because simulations of
   planetary formation indicate a wide range of possible water
   contents \citep{raymond_et_al2004}.  In all cases, the land and
   ocean are uniformly distributed over the planet so that each
   latitude band has the same percentage of land and ocean
   coverage. Due to the lower specific heat capacity of land compared
   to water, altering the land distribution to produce polar
   continents could provide additional stability against snowball
   states.  See \citet{spiegel_et_al2009} for a detailed discussion of
   climate models of planets with nonuniform ocean coverage.}
\end{enumerate}

\section{Model Validation}
\label{sec:modelval}
\citet{spiegel_et_al2008} confirmed that our EBM works reasonably well
for the Earth and reproduces the Earth's climate to a degree of 
accuracy sufficient for investigations of exoplanet habitability.
With the exception of the north/south asymmetry in the Earth's
temperature profile at latitudes south of 60$^\circ$S caused by
Antarctica, the model agrees with the Earth's observed temperatures 
in 2004, as compiled by the National Center for Environmental 
Protection/National Center for Atmospheric Research
\citep{kistler_et_al1999, kalnay_et_al1996}.\footnote{Annually, zonally-averaged temperatures vary little from 
year-to-year.}
\citet{spiegel_et_al2009} verified that the model
predicts the seasonal variations in the Earth's radiative fluxes,
by comparing the model results to data from NASA's Earth Radiation 
Budget Experiment \citep{barkstrom_et_al1990}. 
The model results also agreed with those of \citet{williams+pollard2003}.

As the current study considers eccentricity variations as well as
obliquity variations, we reexamine the model relaxation 
time to ensure that the model run time of 130 years used in 
\citet{spiegel_et_al2008, spiegel_et_al2009} is still sufficient 
for the model to reach a relaxed state under the forcing 
conditions explored here.  
As shown in the Appendix, model runs reach a 
stable oscillatory state
within 100 years of model evolution.  Thus, running the model 
for 130 years ensures that the resulting temperature profile represents 
a relaxed state of the planet, independent of
initial conditions.

As a further check, we compare our results to those of
\citet{williams+pollard2002}.  In that work, a latitudinally resolved
EBM and the three-dimensional climate code GENESIS 2 are used to model
the climate of an Earth-like planet in an eccentric orbit with a
semimajor axis of 1~AU. The geography, atmosphere, and obliquity
angles of their model planet were identical to those of the Earth.
Figure \ref{fig:wpcomp} is our version of their paper's Figure 2: we
conducted model runs of planets with obliquity angles ${\theta_p =
23.5^\circ}$ and ${\theta_a = 0^\circ}$, semimajor axis $a = 1$~AU, and
eccentricities $e$~=~0.1, 0.3, and 0.4.  While our results do not
strictly reproduce those of \citet{williams+pollard2002}, the general
shapes of the temperature curves are similar and the temperatures
generally agree to within 5~K.  This agreement between our EBM and
both the EBM and the general circulation model used by
\citet{williams+pollard2002} gives us further confidence in the
suitability of our EBM for the habitability studies presented below.

\section{Study of Habitability}
\label{sec:results}
We present a suite of models designed to probe the maximum semimajor
axis at which a planet remains habitable before transitioning to a
snowball state. We view habitability as a continuous, rather than a
discrete, property and consider both temporal and regional
habitability \citep{spiegel_et_al2008, spiegel_et_al2009}.  We also
follow convention by adopting the freezing and boiling points of water
under 1~bar of atmospheric pressure as the lower and upper bounds on
habitable temperature.  While boiling temperatures may seem extreme,
there are several hyperthermophiles on Earth that can grow at
temperatures above 373~K \citep[e.g.][]{kashefi+lovley2003,
takai_et_al2008} so even our definition of habitability may
be conservative and Earth-centric. Regardless, for the purpose of
this paper, regions of a planet that are at temperatures between 273~K
and 373~K are considered habitable while regions outside that
temperature range are considered not habitable.  We
note that some regions of a planet's surface may be habitable even
when the rest of the surface is not or that a planet may be habitable
for only part of a year.  Accordingly, we refer to both the temporal
habitability fraction (the fraction of a year for which a given
latitude band is habitable) and the regional habitability fraction
(the fraction of the surface that is habitable at a given time). A
detailed description of these terms is provided in
\citet{spiegel_et_al2008}.

When a planet becomes globally frozen year-round, we say that it has
fallen into a ``snowball'' state.  Here, we examine the maximum
allowed semimajor axis that our models can withstand before falling
into a snowball state.  Recall that the models in this paper do not
include longterm geochemical feedback processes that would tend to
stabilize a geophysically active planet's climate against such a
snowball transition.  As proposed by \citet{walker_et_al1981}, the
decreased efficiency of the carbonate-silicate weathering cycle at low
temperatures should cause greenhouse gases from volcanic eruptions to
accumulate in the atmosphere and gradually warm a planet out of a
near-snowball state.  Models by \citet{kasting_et_al1993} that
incorporated this negative feedback loop showed that including the
effects of the carbonate-silicate cycle extends the outer edge of the
Sun's habitable zone to at least 1.37~AU.  Since our model does not
incorporate this feedback loop, our simulated planets may be more
prone to snowball transitions than actual planets.  Nevertheless,
short-term changes in forcing may induce a snowball transition in far
less time than the $\sim$million year period that would be required to
accumulate enough CO$_2$ in the atmosphere to restore temperate
conditions. The value of our fixed-atmosphere models is to probe
circumstances in which short-term (destabilizing) feedbacks might
induce a snowball ``phase-transition'' that overwhelms longer-term
(stabilizing) feedbacks.

\subsection{Probing the Outer Limits of Habitability}
\label{ssec:OHZ}
We probe the outer limits of habitability with a variety of diagnostic
tests.  We examine the effect of increasing semimajor axis while
holding $e$ constant (Figure \ref{fig:ee06}) and the effect of
increasing $e$ at a constant semimajor axis (Figures \ref{fig:au1},
\ref{fig:thetap0}, and \ref{fig:thetap23}).  We also explore the
relative influence of rotation rate and eccentricity for model planets
at range of semimajor axes (Figure \ref{fig:outerlim}).  Figures
\ref{fig:thetap0}, \ref{fig:thetap23}, and \ref{fig:outerlim} show
dramatically increased outer boundaries of habitability for some
eccentric models.  In Figure \ref{fig:thetap0}, a planet with $e=0.9$
and $\theta_p = 0\degr$ has nonzero habitability out to 2.85~AU; in
Figure \ref{fig:thetap23}, a planet with $e = 0.7$ and
$\theta_p = 23.5\degr$ is partially habitable to 1.215~AU; and in Figure
\ref{fig:outerlim}, a planet with eccentricity of only 0.5 and
$\theta_p = 90\degr$ (perhaps the most likely obliquity) is partially
habitable out to 1.90~AU.  These models all have Earth-composition
atmosphere.

Consider for instance the models in Figure \ref{fig:ee06} for a planet
with eccentricity ${e = 0.6}$, rotation rate ${\Omega_p =
\Omega_\earth}$, obliquity angles ${\theta_p = 23.5^\circ}$ and
${\theta_a = 0^\circ}$, and ocean fraction = 70\%.  For a semimajor
axis of ${1.025}$~AU, the planet is completely habitable throughout
the course of a year, but shows a strong temperature asymmetry due to
the alignment of northern ``winter'' with periastron. Interestingly,
the northern hemisphere actually decreases in temperature during
northern ``summer.''  This occurs because, even though the northern
hemisphere is pointed toward the star, the planet is at apoastron and
receives much less insolation than when at periastron (see Figure
\ref{fig:rel}).  The southern hemisphere faces away from the star at
apoastron and becomes much colder than the northern hemisphere because
it receives even less insolation during its long winter.

If the semimajor axis of the planet is increased to 1.050~AU, the
southern hemisphere becomes so cold during the winter that a permanent
ice cover develops over the southern pole in late southern spring.  If
the semimajor axis is increased to 1.075~AU, the ice forms earlier in
the year because the planet cools faster and spends less time near the
star at periastron.  In addition, the ice that develops on a planet at
$a = 1.075$~AU also extends farther northward before melting at
periastron than the ice formed at smaller semimajor axes.  Once the
semimajor axis reaches 1.100~AU, however, the southern region of the
planet becomes so cold during southern winter and spring that it
cannot warm sufficiently at periastron and remains below freezing
year-round.  In addition, a seasonal northern polar cap develops
during northern winter despite the proximity of the star.  When the
semimajor axis is further increased to 1.125~AU, the magnitude of the
ice-water albedo effect is so strong that the entire planet
transitions to a snowball state because more ice is formed during the
southern winter at apoastron than can be melted during southern summer
at periastron.

As expected from the discussion in Section \ref{sec:milcyc}, the
maximum habitable semimajor axis moves outwards with increasing
eccentricity.  Figure \ref{fig:au1} displays the planetary temperature
at each latitude over the course of one orbit for planets in orbits
with ${a = 1}$~AU and increasing eccentricities.  At low eccentricity
neither pole receives enough insolation to raise the temperature above
freezing during any part of the year, but, as the orbit becomes more
eccentric, the net annual irradiation received by the planet increases
(see Figure \ref{fig:rel}) and the habitable region of the planet
expands.  The northern pole becomes habitable at a lower eccentricity
than the southern pole because the southern hemisphere faces away from
the star at apoastron and therefore absorbs less annually averaged
flux than the northern hemisphere.  The southern hemisphere also
experiences a longer, colder winter than the northern hemisphere,
which faces toward the star at apoastron and away from the star at
periastron.  The model planet's polar climate differs from that of the
Earth because of a variety of simplifications that the model has
compared to the Earth, including uniform distribution of continents
and ocean, the azimuthal obliquity of 0$\degr$ (in comparison to the
Earth's value of 13$\degr$), the various climate feedbacks that our
model does not include, and our treatment of heat redistribution as a
purely diffusive process.

A visual example of the outward movement of the maximum habitable
semimajor axis with increasing eccentricity is provided in the left
panel of Figure \ref{fig:thetap0}, which displays the temporal
habitability fraction of a series of planets with polar obliquity
${\theta_p = 0^\circ}$ and a range of eccentricities and semimajor
axes.  As the eccentricity of the planet's orbit is increased, the
maximum habitable semimajor axis also increases for each latitude band
of the planet.  The most extreme example is shown in the last row of
Figure \ref{fig:thetap0} for planets with ${e = 0.9}$.  At such high
eccentricities, the range of semimajor axes for habitable planets lies
past the maximum habitable semimajor axis for planets in
low-eccentricity orbits (${e \lesssim0.10}$).  In addition, the
variation of habitability with latitude is much more noticeable for
the case of ${e = 0.9}$ planets than for the lower eccentricity
planets. At the equator, this model maintains partial habitability out
to 2.85~AU.  While this is an interesting suggestion that such
highly eccentric planets could have dramatically expanded habitable
zones, this result should be viewed cautiously, since such extremely
eccentric planets might not be reasonably modeled by a simple EBM.

Seasonal variations in habitability are much more pronounced on
planets in highly eccentric orbits.  As shown in the right panel of
Figure \ref{fig:thetap0}, the habitability of a planet in a
near-circular orbit is nearly constant year-round, but the
habitability of planets with ${e \gtrsim0.30}$ depends on the season.
Because the models were initialized at periastron, the planet receives
much more stellar irradiation during the first half of the year (time
$<$ 0.5 years) than during second half of the year (time $>$ 0.5
years).  During the long winter near apoastron, ice accumulates at the
poles and decreases the habitability fraction of the planet, especially
at increased values of the semimajor axis.  Near periastron, however,
the seasonal ice cover melts and the regional habitability fraction is
increased.  Accordingly, the regional habitability fraction for
planets on highly eccentric orbits depends strongly on both the
semimajor axis and the time of the year. 


Since the model planets analyzed in Figure \ref{fig:thetap0}
have ${\theta_p = 0^\circ}$ and ${\theta_a = 0^\circ}$, their temporal
habitability fraction is symmetric with respect to the equator.  For
planets with non-zero $\theta_p$, the northern and southern
hemispheres display different temporal habitabilities.  Figure
\ref{fig:thetap23} shows the temporal and regional habitability
fractions for a set of model planets with $\theta_p = 23.5\degr$ and a
range of eccentricities.  As displayed in the left panel, the northern
hemisphere of such planets is more habitable than the southern
hemisphere at larger semimajor axes.  This effect is most pronounced
for planets in highly eccentric orbits.

The right panel of Figure \ref{fig:thetap23} demonstrates
that planets with eccentricities between 0.4 and 0.6 experience sharp
decreases in regional habitability at the snowball transition.
Just inside the maximum habitable semimajor axis, at least 60\% of the
surface of a planet with a 24-hr day or a 72-hr day is habitable for
the majority of the year. Less than 0.005~AU beyond this distance,
however, the entire planet becomes completely non-habitable.  At
higher eccentricities, the snowball transition is much more
gradual. For eccentricity 0.7, a small region of the planet remains
transiently habitable for semimajor axes between 1.2~AU and 1.225~AU
regardless of rotation rate.  Consequently, the regional habitability
plot reveals a sharp cut-off in regional habitability at the snowball
transition for planets with moderate eccentricities (${0.4 < e <
0.6}$) but a long ``tail'' of decreasing partial habitability for
slightly higher eccentricity (${e = 0.7}$).

Figure \ref{fig:outerlim} shows the semimajor axis corresponding to
the snowball transition for planets with 8-,\\ 24-, and 72-hour
days. The semimajor axis indicated is the largest semimajor axis for
which at least part of the planet is habitable at some point in the
year.  For reference, the black curve plots the ratio of the annual
mean flux received on an orbit with the eccentricity shown along the
x-axis to the flux received on a circular orbit at 1~AU.  As displayed
in the figure, the assumption that the location of the habitable zone
scales with the orbit-averaged flux is justified for low to moderate
eccentricity orbits ({$e < 0.65$}), but for highly eccentric orbits,
the habitable zone extends to much greater distances than that simple
scaling relationship predicts. For {$e = 0.75$}, for example, model
planets with {$\theta_p = 23.5^\circ$}, {$\theta_a = 0$}, and {$f_o =
0.7$} are habitable out to $a\sim$1.39~AU even though the scaling
relationship would predict an outermost habitable semimajor axis of
only $\sim$1.23~AU.  Despite the simplicity of this scaling relation,
our ansatz is typically within the identified snowball transition
region for a planet with a 24-hour day.

Intriguingly, in the context of our models, the relationship between
rotation rate and the position of the snowball transition does not
seem to be monotonic. In a circular orbit, a planet with an 8-hr day,
{$\theta_p = 23.5^\circ$}, {$\theta_a = 0$}, and {$f_o = 0.7$} undergoes
a snowball transition at $a\sim$0.95~AU, but more slowly rotating
planets (24-hr days or 72-hr days) are habitable out to $a\sim$1~AU.
Consequently, more quickly rotating planets in circular orbits freeze
over at shorter distances than more slowly rotating planets.  However,
if the eccentricity of the orbit is raised to $e = 0.2$, planets with
Earth-like 24-hr days are habitable at greater semimajor axes
($a\sim$1.013) than planets with either 8-hr days ($a\sim$1.003) or
72-hr days ($a\sim$0.993). These differences are smaller than the
difference in the position of the snowball transition for quickly
rotating (8-hr days) and less rapidly rotating planets (24-hr or 72-hr
days) in circular orbits, but the order of the distances of the
snowball transitions is unexpected. As discussed by
\citet{spiegel_et_al2008}, this suggests that there may be a trade-off
in keeping the equator warm by rotating sufficiently quickly that not
all of the heat can diffuse to the poles and by rotating slowly enough
to allow enough heat to diffuse to the high latitudes to prevent the
formation of large-scale polar ice coverage that could cool the entire
planet through the ice-water albedo effect. Exploring a denser range
of rotation rates across a variety of eccentricities might help
elucidate the relationship between rotation rate, eccentricity, and
the position of the snowball transition in low eccentricity
orbits. For more eccentric orbits, however, rotation rate (or at least
the rotation rates studied here) does not appear to have a significant
influence on the position of the snowball transition. As shown in
Figure \ref{fig:outerlim}, for orbits with $e = 0.5$, the onset of the
snowball transition occurs at the same semimajor axis for planets with
8-hr, 24-hr, and 72-hr days. Additionally, the position of the
snowball transition is nearly identical for planets with 24-hr and
72-hr days in orbits with e$\gtrsim0.35$.

\subsection{Sensitivity Study}
\label{ssec:sense}
Previous models of habitability \citep{spiegel_et_al2008,
spiegel_et_al2009, williams+pollard2002, williams+pollard2003,
williams+kasting1997} have investigated a variety of test planets, but
there is still a large region of parameter space unexplored.  The
sheer number of factors influencing climatic habitability means that
thousands of model runs would be required to fully explore the
contours of the snowball transition in the multi-dimensional space of
parameters describing the star, the orbit, the planetary spin (rate
and obliquity), and properties of the planet's atmosphere and surface.
Instead, we present the results of a sensitivity study to determine
which parameters have the strongest effects on habitability.  We find
that increasing the polar obliquity increases the semimajor axis of
the snowball transition for both low and high eccentricity planets,
but that the effects of changes in azimuthal obliquity or ocean
coverage depend on eccentricity.

We consider two model planets, one with eccentricity $0.2$ and the
other with eccentricity $0.5$.  Both model planets have Earth-like
obliquity (${\theta_p = 23.5^\circ}$, ${\theta_a = 0^\circ}$ compared
to ${\theta_{p_\earth} = 23.5^\circ}$, ${\theta_{a_\earth} =
13^\circ}$ for the Earth) and uniform continent distributions with an
Earth-like 70\%~ocean fraction.  We first determine the maximum
habitable semimajor axis for both planets by conducting preliminary
model runs on a fine grid in semimajor axis (0.005~AU spacing).  Then,
we systematically alter each parameter either individually or in
combination to determine the maximum semimajor axis for a habitable
planet as a function of slight deviations in each input parameter.
Our results for the planet with ${e = 0.2}$ are shown in Table
\ref{tab:sense02} and our results for the planet with ${e = 0.5}$ are
shown in Table \ref{tab:sense05}.

For both cases increasing the polar obliquity allows the planet to
remain habitable at greater semimajor axes.  When ${\theta_p
\lesssim57^\circ}$, the equator receives more annually averaged
insolation than the poles and the direction of heatflow is poleward,
but at higher polar obliquities, the poles receive more insolation
than the equator and the direction of heatflow is reversed.  As a
result, while planets with ${\theta_p = 0^\circ}$ develop permanent
ice coverage at both poles and transition to a snowball state when the
downward extension of the ice reaches the equator, planets with
${\theta_p = 90^\circ}$ are warmest at their poles and heat is
transported from the poles toward the equator.  Planets with
${90^\circ}$ polar obliquity therefore develop small, seasonal ice
coverage near the equator and remain habitable for an additional
0.04~AU (for ${e = 0.2}$) or for an additional 0.838~AU (for ${e =
0.5}$). This later case is the planet that is habitable at 1.90~AU
featured in Figure \ref{fig:outerlim}, which is transiently habitable
at the south pole during the brief, intense periastron summer.

Reducing the ocean fraction from 70\% ocean to 10\% ocean decreases
the maximum habitable semimajor axis for planets with high
eccentricity, but has a more complicated effect on planets with low
eccentricity depending on polar obliquity. Consider, for example, a
desert planet with ${\theta_p = 90^\circ}$, ${a = 1.255}$~A, and ${e =
0.2}$. The desert planet experiences tremendous temperature
oscillations during the course of the year, but the southern pole
becomes transiently habitable during northern winter at periastron.
The southern pole then freezes during the long winter and reaches cold
temperatures (${T \sim}$150~K) before thawing again at apoastron.  As
shown in Figure \ref{fig:ee20verydryob90}, the southern pole is the
only part of the planet that is ever within the liquid water limits of
habitability and experiences the most extreme temperature variations
on the planet.  However, properly modeling this planet would require
taking into account the latent heat of freezing and melting.

Variations in azimuthal obliquity shift the positions of the solstices
with respect to periastron and apoastron.  As a result, the azimuthal
obliquity can influence the orbital distance at which a planet
transitions to a snowball state. At low eccentricity, there is not
much effect.  At higher eccentricity, $\theta_a\sim 30 \degr$ leads to
a larger habitable semimajor axis than ${\theta_a = 0\degr}$ or
${\theta_a = 90\degr}$. Why might this occur? At low azimuthal
obliquity, the contrast between the annually averaged flux received by
the southern and northern hemispheres is conducive to ice growth.
Conversely, for $\theta_a\sim 90 \degr$ the planet experiences milder
winters, but the less intense summers reduce seasonal melting. A sweet
spot occurs near $\theta_a = 30\degr$, where the enhanced irradiation
experienced during summer roughly compensates for the colder winters.
As shown in Figure~\ref{fig:rel}, the ratio of the flux received at
periastron to the flux received at apoastron increases with increasing
eccentricity so that variations in azimuthal obliquity have a more
pronounced effect on planets in more eccentric orbits.

\section{Summary and Conclusion}
\label{sec:conc}
We presented the results of a series of idealized energy balance model
runs to determine the habitability of planets with a range of
eccentricities.  As shown in Section \ref{ssec:OHZ}, planets in orbits
with a given semimajor axis can remain habitable for a range of
eccentricities.  This suggests that if a planet were to experience
eccentricity perturbations caused by a giant planet companion, as
considered in the companion paper \citep{spiegel_et_al2010b}, it would
not necessarily transition to a snowball state.  Instead, our study
suggests that many perturbed planets could remain fully or partially
habitable even at slightly different eccentricities.  Intriguingly, a
previously frozen planet might thaw if perturbed to a higher
eccentricity.  Incorporating the latent heat involved in the melting
of a frozen world is a complex problem that we do not address in this
paper, but we propose a solution in \citet{spiegel_et_al2010b}.

Throughout this study, we have used the conventional liquid water
definition of habitability, but many extremophiles are capable of
surviving outside of that temperature range
\citep{carpenter_et_al2000, kashefi+lovley2003, takai_et_al2008,
junge_et_al2004}.  While it remains to be seen whether life could
originate at temperatures well below 273~K or well above 373~K, we
should remain cautious about making any assumptions about the
limitations of microbes.  Recent advances in biology continue to
demonstrate that lifeforms on Earth are far more inventive and
ubiquitous than we ever would have expected.

In this study we have considered the effects of eccentricity on the
semimajor axis corresponding to the snowball transition for
pseudo-Earth planets and found, as has been seen by
\citet{williams+pollard2002}, that increasing eccentricity can
increase the allowed semimajor axis for habitable planets.  In
addition to increasing the maximum semimajor axis at which the planet
can remain habitable, increases in eccentricity also enhance regional
and seasonal variability in planetary temperatures and lead to a more
gradual transition from habitable to non-habitable planets with
increasing semimajor axis.

We have also conducted a sensitivity study to determine which orbital
parameters are the most influential on the location of the snowball
transition.  Although changing the obliquity of a low-eccentricity
planet can alter the maximum habitable semimajor axis by a hundredth
of an AU, we find that reducing the ocean fraction has the strongest
effect on the maximum habitable semimajor axis of a low-eccentricity
planet because of the important role of the ocean's thermal inertia in
mediating climate variations.  Combining decreases in ocean coverage
with increases in polar obliquity can further extend the position of
the snowball transition, but changes in azimuthal obliquity have a
negligible effect on the habitability of low eccentricity planets.

The situation for higher-eccentricity planets is more complicated, but
variations in polar obliquity seem to have the most powerful effect on
the position of the maximum habitable semimajor axis and can increase
the semimajor axis corresponding to the snowball transition by
$\sim$0.8~AU. Changes in azimuthal obliquity are also significant for
highly eccentric planets because of the uneven distribution of flux
throughout the orbit.  Increasing the azimuthal obliquity to
${90^\circ}$ actually decreases the semimajor axis of the snowball
transition, but there is a sweet spot near $\sim$30$^\circ$ where the
increase in the azimuthal obliquity extends the position of the
snowball transition by $\sim$0.175~AU.  Our simulations indicate that
changes in the effective thermal diffusivity by roughly an order of
magnitude in either direction (motivated by the suggestion that
thermal diffusivity might depend strongly on rotation rate) have
little influence on habitability for planets with moderate
eccentricity (${0.35 \lesssim e \lesssim 0.65}$), but planets with
high or low eccentricity display a complex dependence of the position
of the snowball transition on diffusivity.

Finally, this study suggests that planets in moderately- or
highly-eccentric orbits (${e \gtrsim 0.5}$) may be habitable to much
larger semimajor axes than would result from simply scaling the
semimajor axis to match the flux received in a circular orbit. In
particular, a model with $e = 0.9$ and $\theta_p =0 \degr$ is
habitable to 2.85~AU and a model with $e = 0.5$ and $\theta_p =
90\degr$ is habitable to 1.90~AU, both with Earth-composition
atmospheres.  Although these numbers are surprisingly large for a
fixed-composition atmosphere, the fact that EBMs tend to be more
susceptible to global freezing than are actual planets gives us some
confidence that our models are not prone to overestimating the outer
boundary of habitability.  The partially habitable model with $e =
0.5$ and $a = 1.90$~AU has apoastron separation of 2.85~AU, which
raises the intriguing possibility that some moderately-to-highly
eccentric planets in the outer reaches of a habitable zone might be
significantly easier to observe with direct imaging platforms such as
TPF-Darwin \citep{kaltenegger+fridlund2005, heap_et_al2008} than
similar planets on circular orbits. Nevertheless, this question cannot
be properly evaluated without a model that appropriately accounts for
both the latent heat of melting/freezing and the atmospheric changes
(in composition, cloud-cover, etc.) that occur with such strong
changes in stellar irradiation over the annual cycle. Ideally, future
studies would combine study of changes in polar obliquity, azimuthal
obliquity, rotation rate, and eccentricity with variations in other
parameters such as continent distribution, atmospheric composition,
and azimuthal obliquity to further explore the multi-dimensional
contours of the snowball transition.

\acknowledgments We acknowledge conversations with Ed Turner, Adam
Burrows, and Lawrence Larmore.  Models were run on the Della
supercomputer at the TIGRESS High Performance Computing and
Visualization Center at Princeton University.  This research was
supported in part by the National Science Foundation under grant
No. PHY05-51164 and by NASA under grant No. NNX07AH68G.  
DSS would like to thank the participants of the ``Revisiting the
Habitable Zone'' meeting and acknowledges support from NASA grant
NNX07AG80G and from JPL/Spitzer Agreements 1328092, 1348668, and
1312647. SNR acknowledges funding from NASA Astrobiology Institutes'
Virtual Planetary Laboratory lead team, supported by NASA under
Cooperative Agreement No. NNH05ZDA001C.

\begin{appendix}
\label{sec:app}
A detailed model relaxation study was completed in
\citet{spiegel_et_al2008}, but because the present work focuses on
variations in eccentricity, we repeat the test to ensure that 130
years of model evolution remains sufficient for the planets in this
study.  Figure \ref{fig:conv} shows the mean global temperature of
several model planets smoothed with a 1-year boxcar filter.  Two of
the model planets have rotation rate $\Omega_p = \Omega_\earth$ and
the third model has rotation rate $\Omega_p = (1/3)\Omega_\earth$. In
this case, the more slowly rotating planet reaches a steady climate
state after a shorter model runtime than the planets with 24-hr
days. However, as shown in detail in the right panel, all of the
models achieve a steady climate state within 100 years and then show
maximum temperature deviations of less than 0.04~K. Specifically, the
model planet with $\Omega_p = \Omega_\earth$, $e = 0.15$, and $a =
1.005$~AU reaches a steady climate state after $\sim$17 years and the
model planet with $\Omega_p = \Omega_\earth$, $e = 0.75$, and $a =
1.125$~AU reaches a steady climate state after $\sim$90 years. The
model planet with $\Omega_p = (1/3)\Omega_\earth$, $e = 0.15$, and $a
= 0.950$~AU achieves a steady climate state after only $\sim$5 years
of model evolution.  The results of this study indicate that
integrating for 130 years is sufficient for this purpose and that the
time required to reach a steady climate state does not increase
monotonically with increasing eccentricity.

\renewcommand{\thefigure}{\Alph{figure}}

\begin{figure}[h]
\centering
\includegraphics[width=0.9\textwidth]{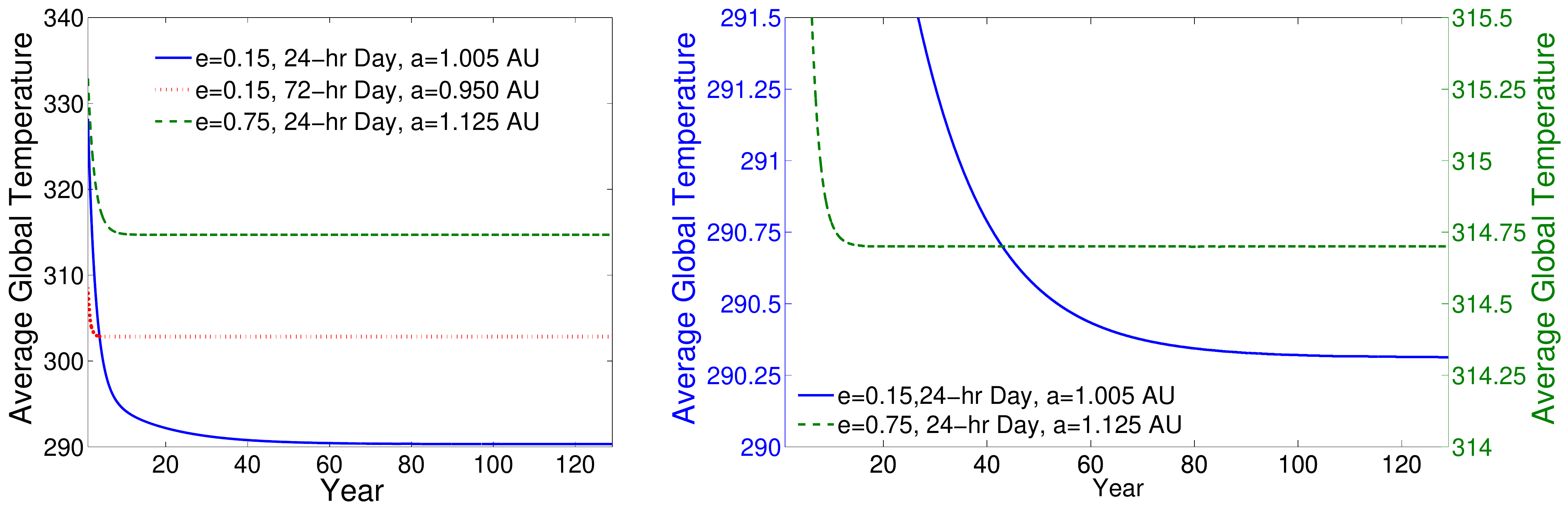}
\caption{Model relaxation tests. {\bf Left:} Variations in the average
  global temperature over the 130-year model evolution for model
  planets with polar obliquity $\theta_p = 23.5^\circ$ and azimuthal
  obliquity $\theta_a = 0^\circ$ and a variety of eccentricities,
  rotation rates, and semimajor axes. The average global temperature
  is smoothed by a 1-year boxcar filter for the entire 130-year model
  evolution of a planet.  The slowly rotating planet reaches a steady
  climate state after $\lesssim$10 years, but as shown in the right
  panel, the planets with 24-hr days require longer model runs to
  reach a steady climate state.  {\bf Right:} A more detailed plot of
  the average global temperature for the model planets with rotation
  rate $\Omega_p = \Omega_\earth$ highlighting the transition to a
  steady climate state. The planet with eccentricity $e = 0.15$ cools
  to $T \sim$290.3~K after $\sim$90 years, but the planet with
  eccentricity $e = 0.75$ reaches a steady climate state with $T =
  314.7$~K after only $\sim$17 years. As before, the global mean
  temperature is smoothed by a 1-year boxcar filter.}
\label{fig:conv}
\end{figure}

\end{appendix}
\newpage

\begin{table}[p]
		\centering
		\caption{Model Values for the Surface Heat Capacity $C$}
		\begin{tabular}{l l }
			\hline \hline
			Surface Type & Effective Heat Capacity (erg cm$^{-2}$ K$^{-1}$)\\ [0.5ex]
			\hline
			Land  & $C_l$ = 5.25 $\times$ 10$^9$ \\
			Ocean & $C_o$ = 40$C_l$ \\
			Ice (263 K $<$ T $<$ 273 K) & $C_i$ = 9.2$C_l$ \\
			Ice (T $<$ 263 K) & $C_i$ = 2.0$C_l$ \\
			\hline
		\end{tabular}
	\label{tab:heatcap}
\end{table}

\begin{table}[p]
		\centering
		\caption{Sensitivity Study for a
		Planet with Eccentricity $e= 0.2$}
		\begin{tabular}{l l l c c}
			\hline \hline
			\multicolumn{3}{c}{Variation in Model} & Outer Edge & Change from \\
			\multicolumn{3}{c}{Parameters} &of HZ (AU) & Fiducial Model (AU)\\
 [0.5ex]
			\hline
			\multicolumn{3}{c}{Fiducial Model (Earth with $e$ = 0.2)\tablenotemark{a}} & 1.010-1.015 & N/A    \\
			$\theta_p$ = 23.5$^\circ$ & $\theta_a$ = 30$^\circ$ & Ocean Fraction = 0.7 & 1.010-1.015 & 0.000  \\
			$\theta_p$ = 23.5$^\circ$ & $\theta_a$ = 90$^\circ$ & Ocean Fraction = 0.7 & 1.010-1.015 & 0.000  \\
			$\theta_p$ = 0$^\circ$    & $\theta_a$ = 0$^\circ$  & Ocean Fraction = 0.7 & 1.015-1.020 & 0.005  \\
			$\theta_p$ = 15$^\circ$   & $\theta_a$ = 0$^\circ$  & Ocean Fraction = 0.7 & 1.010-1.015 & 0.000  \\
			$\theta_p$ = 30$^\circ$   & $\theta_a$ = 0$^\circ$  & Ocean Fraction = 0.7 & 1.010-1.015 & 0.000  \\
			$\theta_p$ = 60$^\circ$   & $\theta_a$ = 0$^\circ$  & Ocean Fraction = 0.7 & 1.030-1.035 & 0.020  \\
			$\theta_p$ = 90$^\circ$   & $\theta_a$ = 0$^\circ$  & Ocean Fraction = 0.7 & 1.050-1.055 & 0.040  \\
			$\theta_p$ = 23.5$^\circ$ & $\theta_a$ = 0$^\circ$  & Ocean Fraction = 0.9 & 1.010-1.015 & 0.000  \\
			$\theta_p$ = 23.5$^\circ$ & $\theta_a$ = 0$^\circ$  & Ocean Fraction = 0.5 & 1.010-1.015 & 0.000  \\
			$\theta_p$ = 23.5$^\circ$ & $\theta_a$ = 0$^\circ$  & Ocean Fraction = 0.1 & 0.990-0.995 & -0.020 \\
			$\theta_p$ = 0$^\circ$    & $\theta_a$ = 0$^\circ$  & Ocean Fraction = 0.1 & 0.995-1.000 & -0.015 \\
			$\theta_p$ = 90$^\circ$   & $\theta_a$ = 0$^\circ$  & Ocean Fraction = 0.1 & 1.255-1.260 & 0.245  \\
\hline
		\end{tabular}
                \tablenotetext{a}{\protect\centering The parameters
                for this planet are $f_o = 0.70$, ${\theta_p = 23.5}$,
                ${\theta_a = 0}$, and ${\Omega_p = \Omega_\earth}$.}
	\label{tab:sense02}
\end{table}

\begin{table}[p]
		\centering
		\caption{Sensitivity Study for a
		Planet with Eccentricity ${e = 0.5}$}
		\begin{tabular}{l l l  c c}
			\hline \hline
		        \multicolumn{3}{c}{Variation in} & Outer Edge & Change from \\
			\multicolumn{3}{c}{Model Parameters} &of HZ (AU) & Fiducial Model (AU)\\
 [0.5ex]
			\hline
			\multicolumn{3}{c}{Fiducial Model (Earth with $e = 0.5$)\tablenotemark{b}}  & 1.070-1.075 & N/A    \\
			$\theta_p$ = 23.5$^\circ$ & $\theta_a$ = 30$^\circ$ & Ocean Fraction = 0.7  & 1.245-1.250 & 0.175  \\
			$\theta_p$ = 23.5$^\circ$ & $\theta_a$ = 90$^\circ$ & Ocean Fraction = 0.7  & 1.065-1.070 & -0.005 \\
			$\theta_p$ = 0$^\circ$    & $\theta_a$ = 0$^\circ$  & Ocean Fraction = 0.7  & 1.070-1.075 & 0.000  \\
			$\theta_p$ = 15$^\circ$   & $\theta_a$ = 0$^\circ$  & Ocean Fraction = 0.7  & 1.070-1.075 & 0.000  \\
			$\theta_p$ = 30$^\circ$   & $\theta_a$ = 0$^\circ$  & Ocean Fraction = 0.7  & 1.365-1.370 & 0.295  \\
			$\theta_p$ = 60$^\circ$   & $\theta_a$ = 0$^\circ$  & Ocean Fraction = 0.7  & 1.765-1.785 & 0.703  \\
			$\theta_p$ = 90$^\circ$   & $\theta_a$ = 0$^\circ$  & Ocean Fraction = 0.7  & 1.900-1.920 & 0.838  \\
			$\theta_p$ = 23.5$^\circ$ & $\theta_a$ = 0$^\circ$  & Ocean Fraction = 0.9  & 1.075-1.080 & 0.005  \\
		        $\theta_p$ = 23.5$^\circ$ & $\theta_a$ = 0$^\circ$  & Ocean Fraction = 0.3  & 1.090-1.095 & 0.020  \\
			$\theta_p$ = 23.5$^\circ$ & $\theta_a$ = 0$^\circ$  & Ocean Fraction = 0.1  & 1.255-1.260 & 0.185  \\
		        $\theta_p$ = 0$^\circ$    & $\theta_a$ = 0$^\circ$  & Ocean Fraction = 0.1  & 1.070-1.075 & 0.000  \\
		        $\theta_p$ = 90$^\circ$   & $\theta_a$ = 0$^\circ$  & Ocean Fraction = 0.1  & 1.415-1.420 & 0.345  \\
			\hline
		\end{tabular}
                \tablenotetext{b}{\protect\centering The parameters
                for this planet are $f_o = 0.70$, ${\theta_p = 23.5}$,
                ${\theta_a = 0}$, and ${\Omega_p = \Omega_\earth}$.}
          	\label{tab:sense05}
\end{table}

\clearpage
\newpage
\bibliography{biblio}
\newpage

\begin{figure}
\centering
\plotone{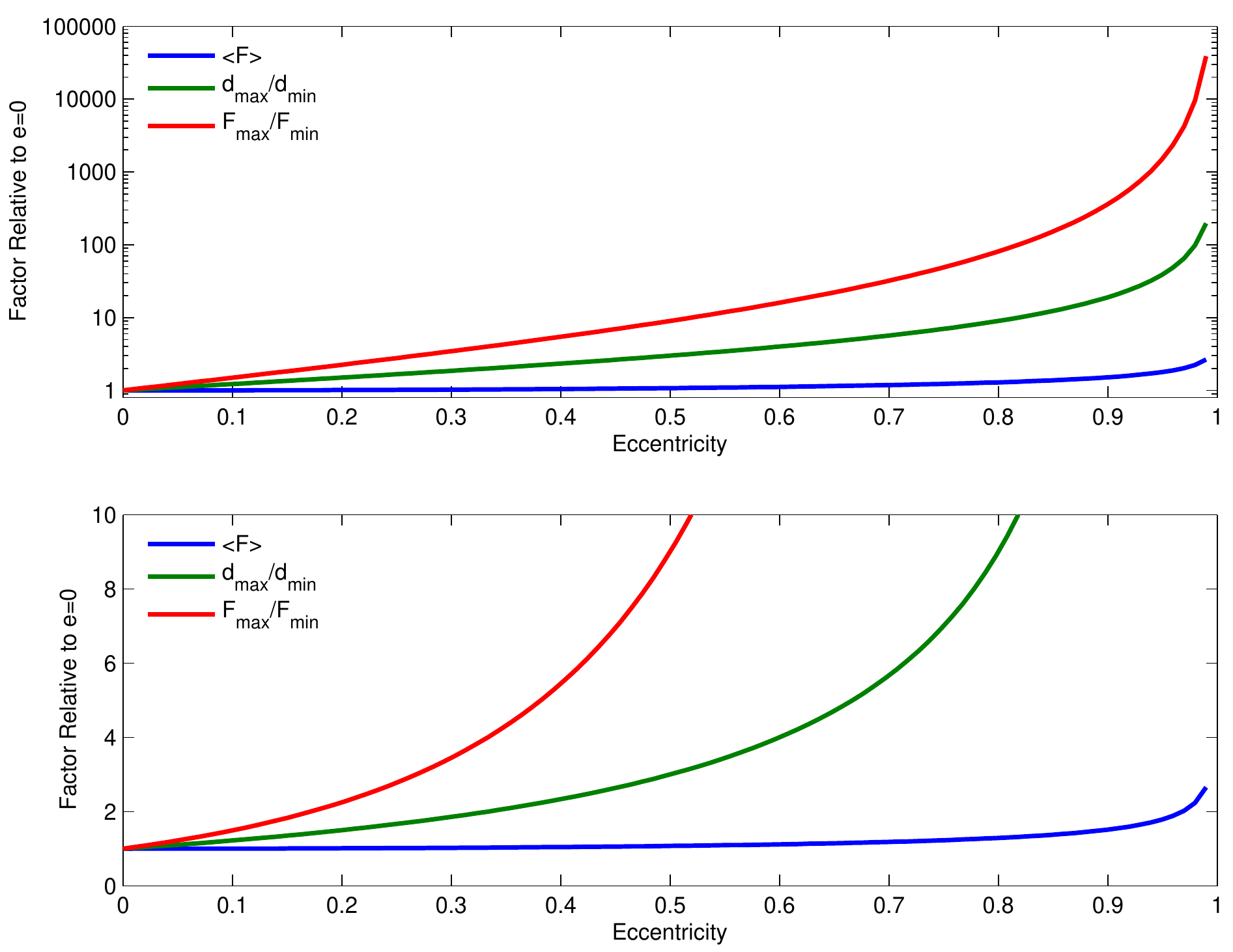}
\caption{Comparison of the mean flux (blue) and the ratio of
periastron flux to apoastron flux (red) for planets in eccentric
orbits to those of planets in circular orbits.  The top plot is
logarithmic and extends from $e = 0$ to $e = 1$ while the bottom plot
is linear to highlight the change in flux between $e = 0$ and $e =
0.5$.  For reference, the green line shows the distance of the planet
at apoastron relative to the distance of the planet at periastron.
Although the average flux changes by less than a factor of two even at
very high eccentricity, the change in flux over the course of the
orbit is several orders of magnitude larger for planets at high
eccentricity.  Consequently, regions of those planets may be both well
below 273~K and well above 373~K during the course of a year.
Provided the planets do not freeze over completely during the long
winter and enter a snowball state, this raises the question of whether
life would be able to withstand such extreme temperature variations.
}
\label{fig:rel}
\end{figure}

\begin{figure}
\centering
\includegraphics[height=1.2\textwidth, angle=90]{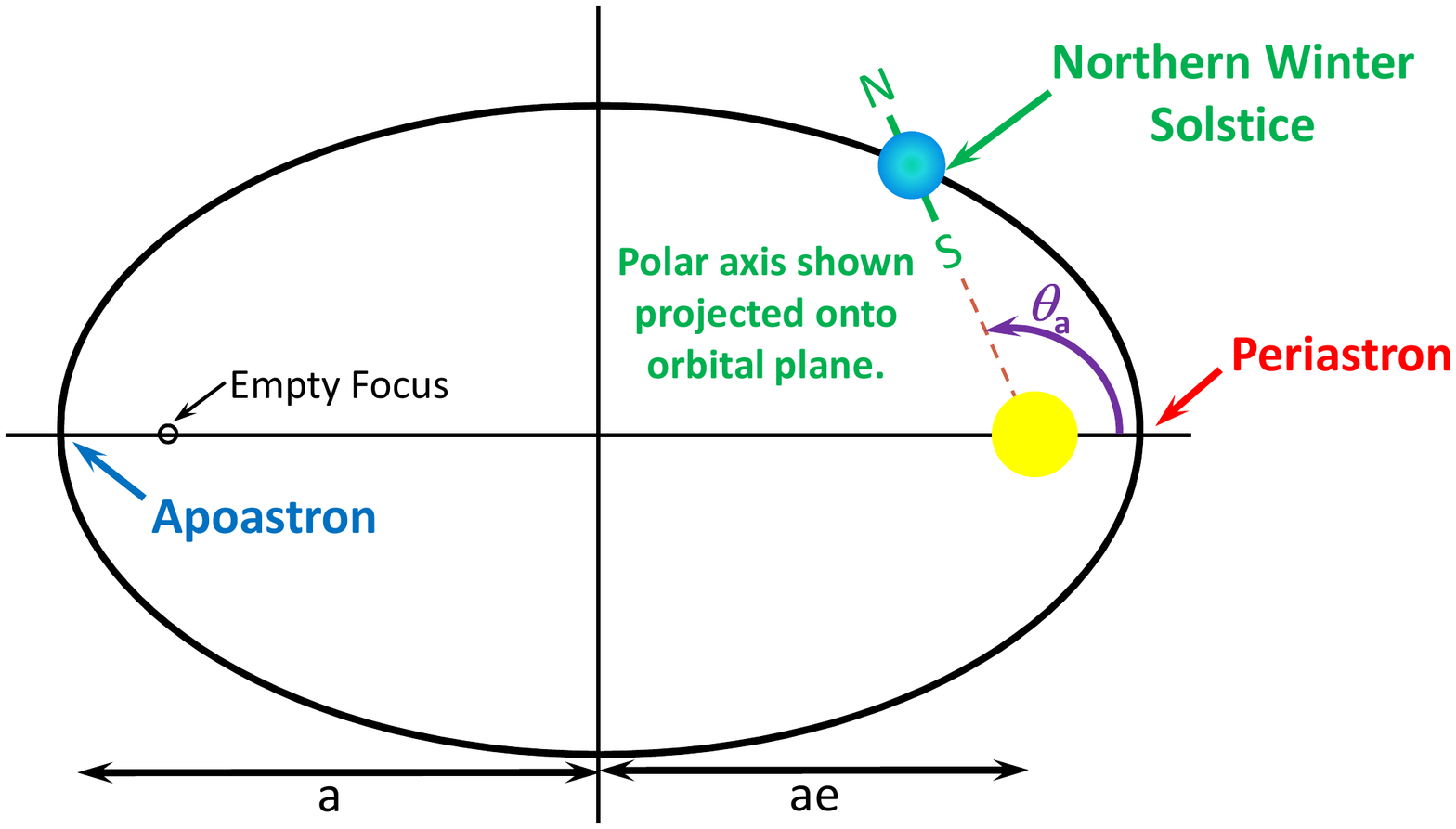}
\caption{Top view of a planet in an elliptical orbit with semimajor
axis $a$ and eccentricity $e$.  As explained in Section
\ref{sec:modelset}, the azimuthal obliquity $\theta_a$ is zero if
periastron is aligned with the projection of the planet's spin angular
momentum vector onto the orbital plane.  In that case, northern winter
coincides with periastron. }
\label{fig:orbparam}
\end{figure}

\begin{figure}
\centering
\includegraphics[width=0.8\textwidth]{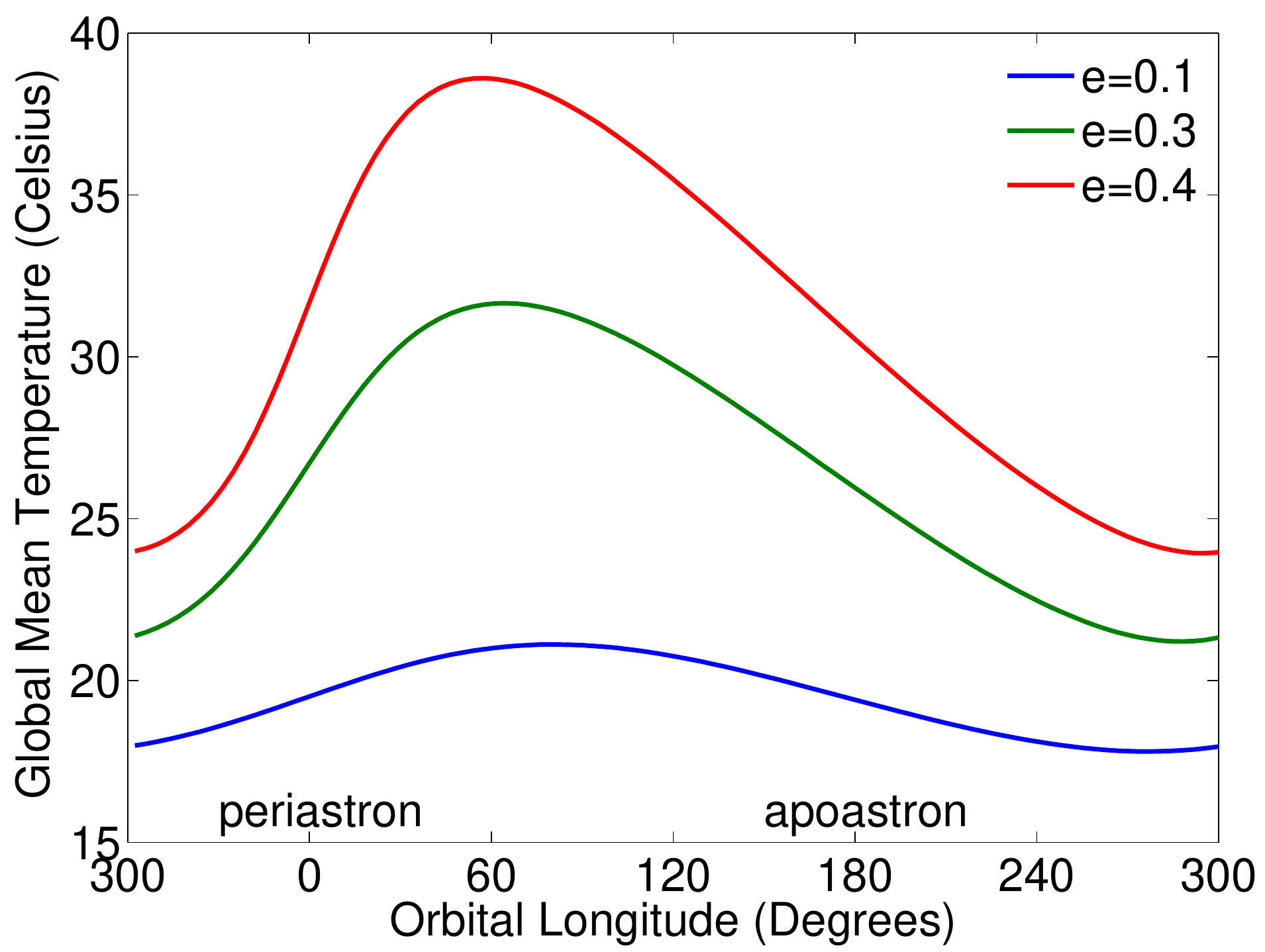}
\caption{Our version of Figure 2 of
  \citet{williams+pollard2002}. Their model used a test planet with the
geography of the Earth while our model used a test planet with a
uniform continent distribution but the same 70\% ocean fraction as the
Earth.  The slight differences between their results and ours might be
due to the presence of an asymmetric southern continent (Antarctica)
in their model but not ours, and other model differences between their
three-dimensional GCM and our simpler one-dimensional EBM.}
\label{fig:wpcomp}
\end{figure}

\begin{figure}[htp]
\centering
\includegraphics[width=0.8\textwidth]{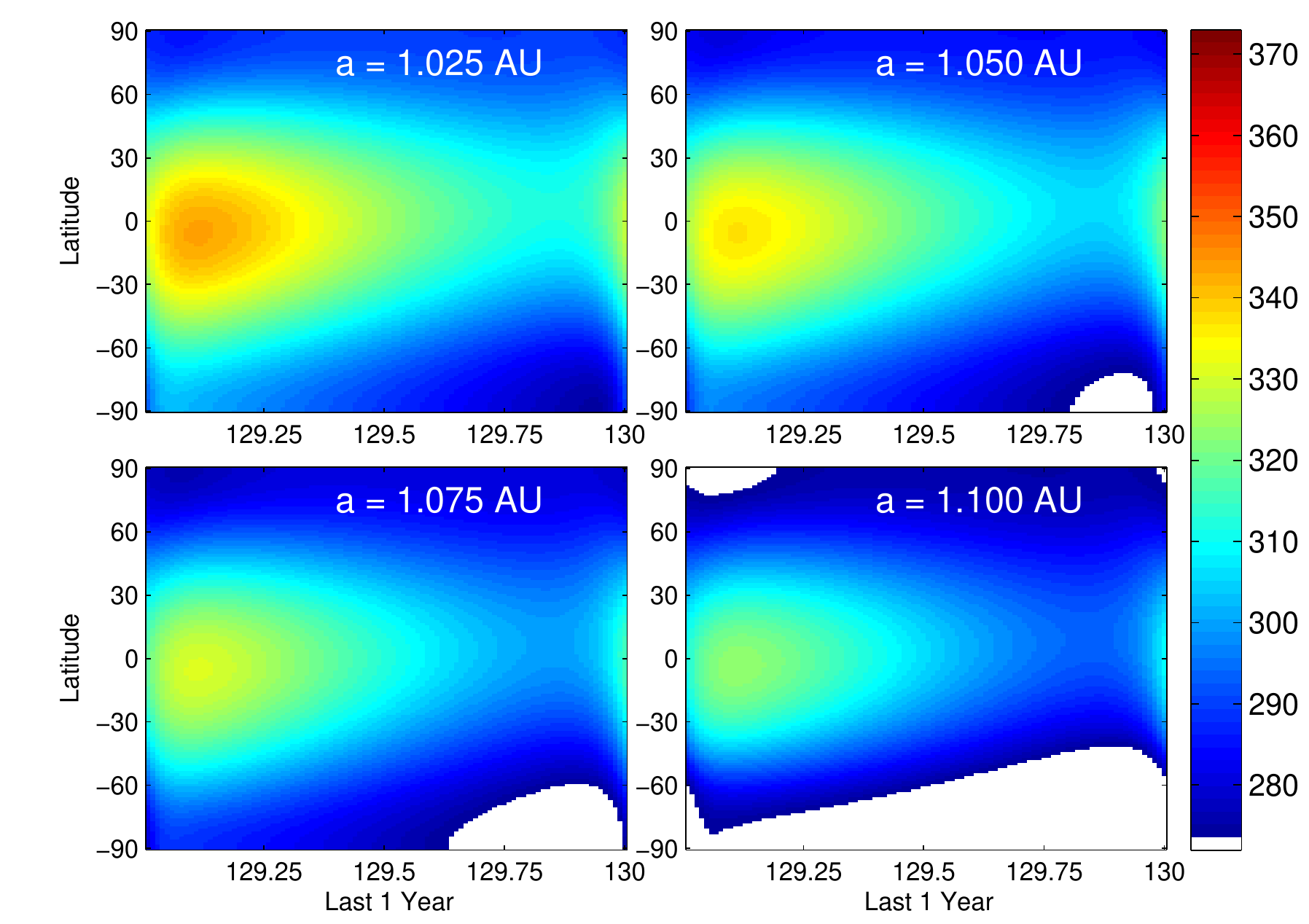}
\caption{Temperature as a function of latitude and time of year for a
  set of planets with $e = 0.6$, $\theta_p = 23.5^\circ$, $\theta_a =
  0^\circ$, $f_o = 70\%$ and rotation rate $\Omega_p = \Omega_\earth$
  with a range of semimajor axes.  The color indicates temperature as
  shown in the colorbar.  Areas of the planet shown in white are
  either below 273 K or above 373 K.  The planet is completely
  habitable at 1.025~AU, but the temperature near the south pole drops
  below freezing during the southern winter at 1.050~AU.  The southern
  ice coverage grows for semimajor axes $> 1.075$~AU and is
  accompanied by northern ice coverage during northern winter at
  1.100~AU.  Beyond 1.125~AU, the planet is completely frozen.}
\label{fig:ee06}
\end{figure}

\begin{figure}
\centering
\includegraphics[width=0.8\textwidth]{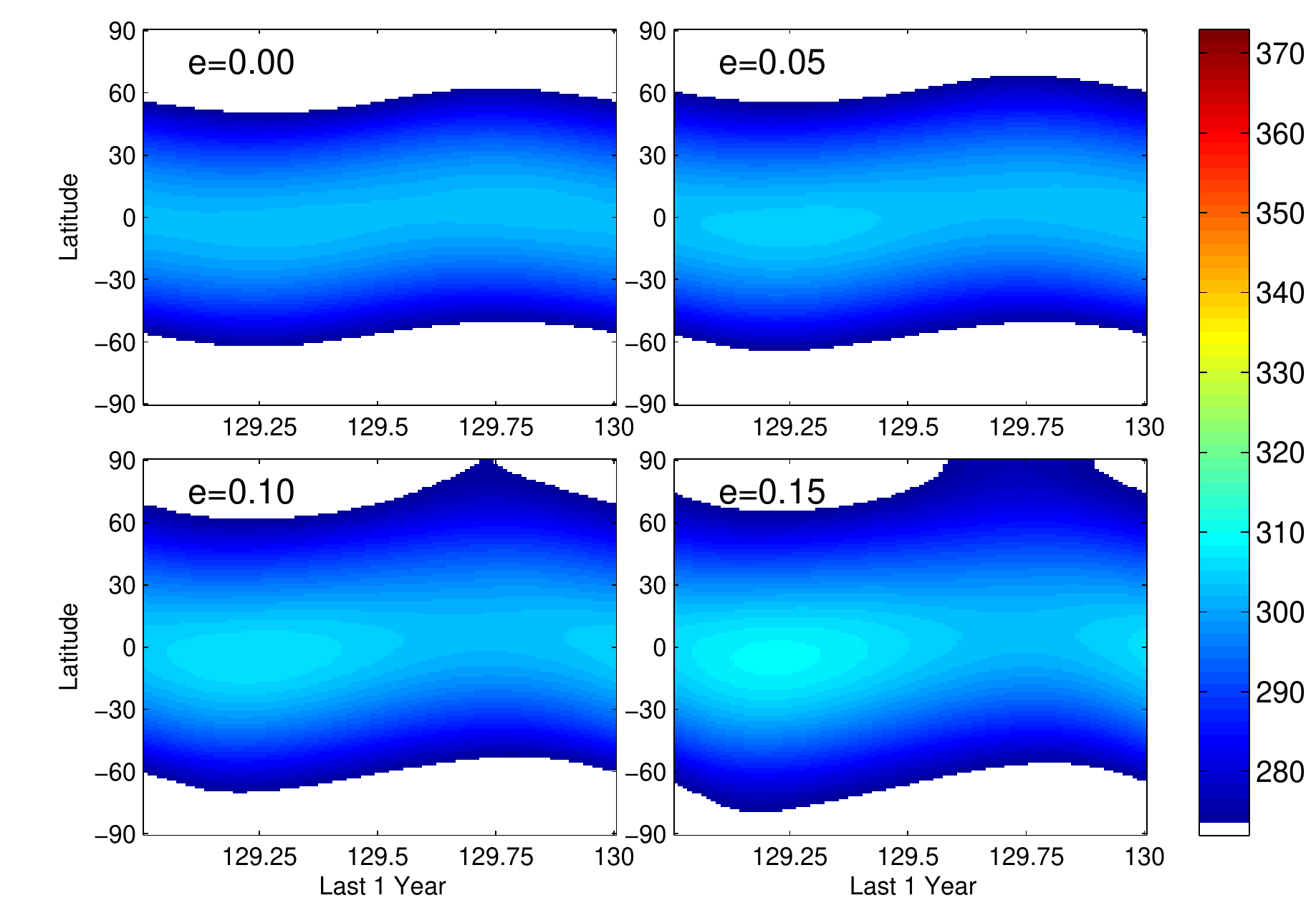}
\caption{Temperature as a function of latitude and time of year for a
  set of planets with $\theta_p$ = 23.5$^\circ$, $\theta_a$ =
  0$^\circ$, $f_o = 70\%$ and rotation rate $\Omega_p =
  \Omega_\earth$.  All of the planets are in an orbit with semimajor
  axis $a = 1$~AU, but the eccentricity varies from $e = 0$ (top left)
  to $e = 0.15$ (bottom right).  At near-zero eccentricity, only the
  mid-latitudes are habitable, but at high eccentricities, the polar
  regions receive greater mean isolation over the course of the year
  and the regional habitability fraction increases. In addition, note
  that the temperature asymmetry between the northern and southern
  hemispheres due to the alignment of periastron with northern winter
  is more pronounced at higher eccentricities because northern winter
  occurs much closer to the star than southern winter.}
\label{fig:au1}
\end{figure}

\begin{figure}
\centering
\includegraphics[width=0.9\textwidth]{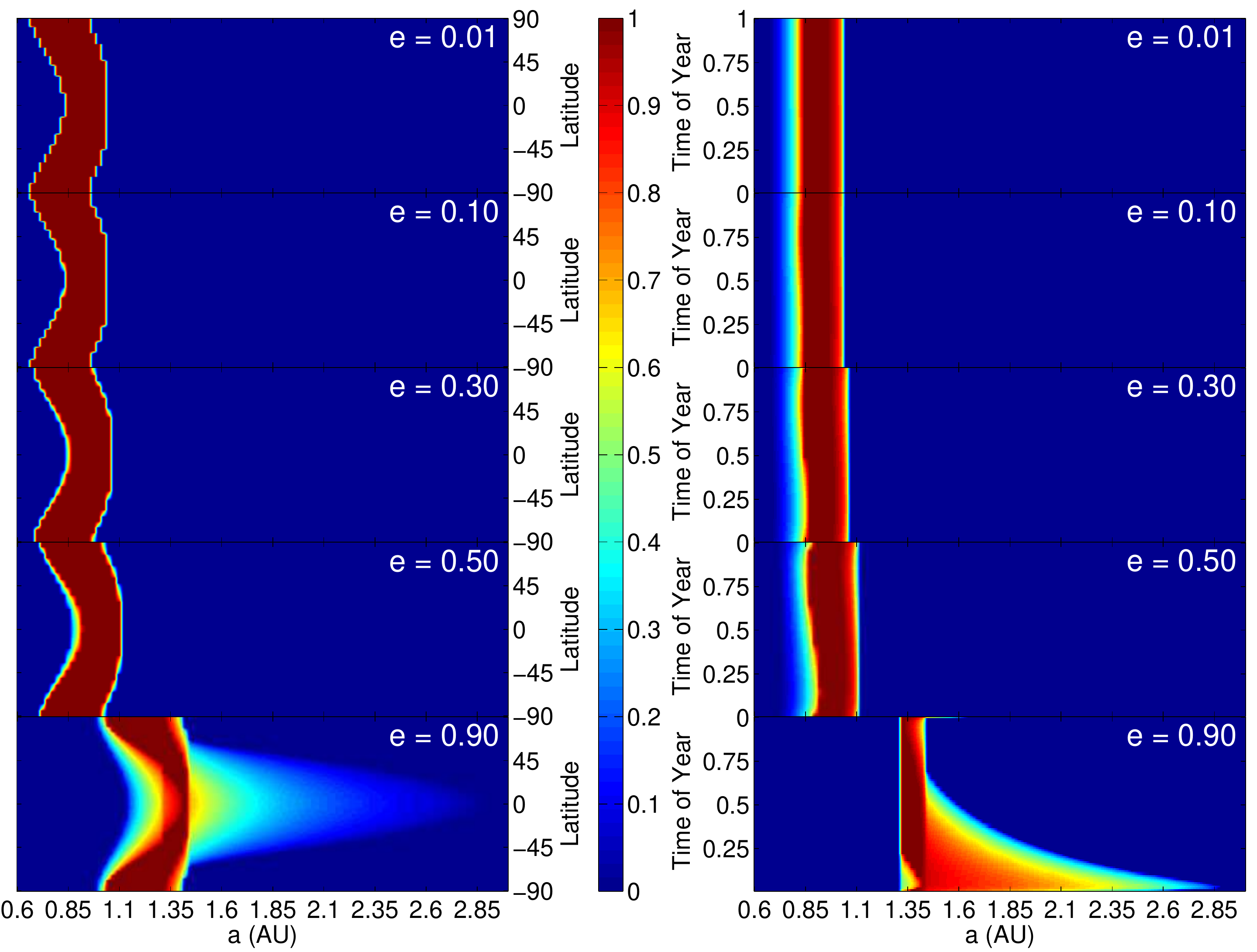}\\
\caption{{\bf Left:} Temporal habitability fraction (the fraction of a
  year for which a given latitude band is habitable) for planets with
  $\theta_p = 0^\circ$, $\theta_a = 0^\circ$, ocean fraction = 70\%,
  and $\Omega_p = \Omega_\earth$.  The eccentricity of each set of
  planets is indicated and increases downward from $e$ = 0.01 to $e$ =
  0.90.  Note that the maximum habitable semimajor axis increases for
  greater eccentricities.  Regions shown in red indicate that the
  region is habitable for the entire year while regions shown in blue
  indicate that the region is non-habitable for the entire
  year. Colors between red and blue signify that the region is
  habitable for part of the year.  {\bf Right:} Regional habitability
  fraction (the fraction of the surface that is habitable at a given
  time) for the same planets as in the left panel.  Note the increased
  seasonal dependence at higher eccentricities due to the extreme
  variations in flux over the orbit.  Times shown in red indicate that
  the entire planet is habitable at that time while times shown in
  blue indicate that the entire planet is non-habitable at that time.
  Colors between red and blue signify that a fraction of the planet is
  habitable at that time. }
\label{fig:thetap0}
\end{figure}

\begin{figure}
\centering
\includegraphics[width=0.8\textwidth]{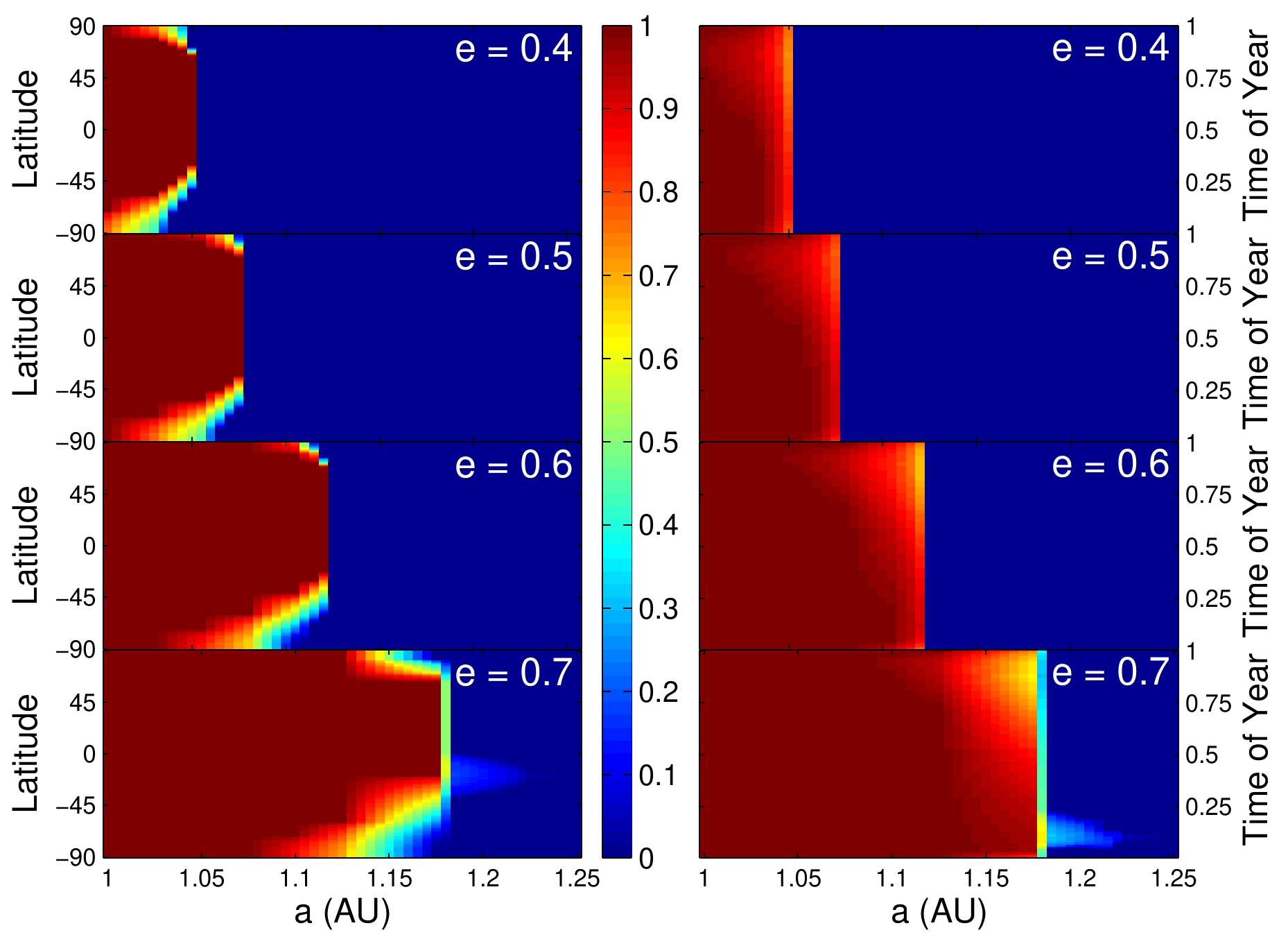}
\caption{ Same as Figure \ref{fig:thetap0} but for planets with
  $\theta_p = 23.5^\circ$. Eccentricity increases downward from $e =
  0.4$ for the top plots to $e = 0.7$ for the bottom plots.
  \textbf{Left:} Temporal habitability fraction as a function of
  semimajor axis. \textbf{Right:} Regional habitability fraction as a
  function of semimajor axis. Note that the light shades of blue at $a
  = 1.2$~AU in the plots corresponding to planets with eccentricity $e
  = 0.7$ means that a small fraction of the planet is habitable for a
  brief period in during southern fall.  Comparison with the left
  panel reveals that the area of the planet that is is transiently
  habitable during southern fall is near $-30^\circ$S, which is near
  the substellar point at periastron. }
\label{fig:thetap23}
\end{figure}

\begin{figure}
\centering
\includegraphics[width=0.8\textwidth]{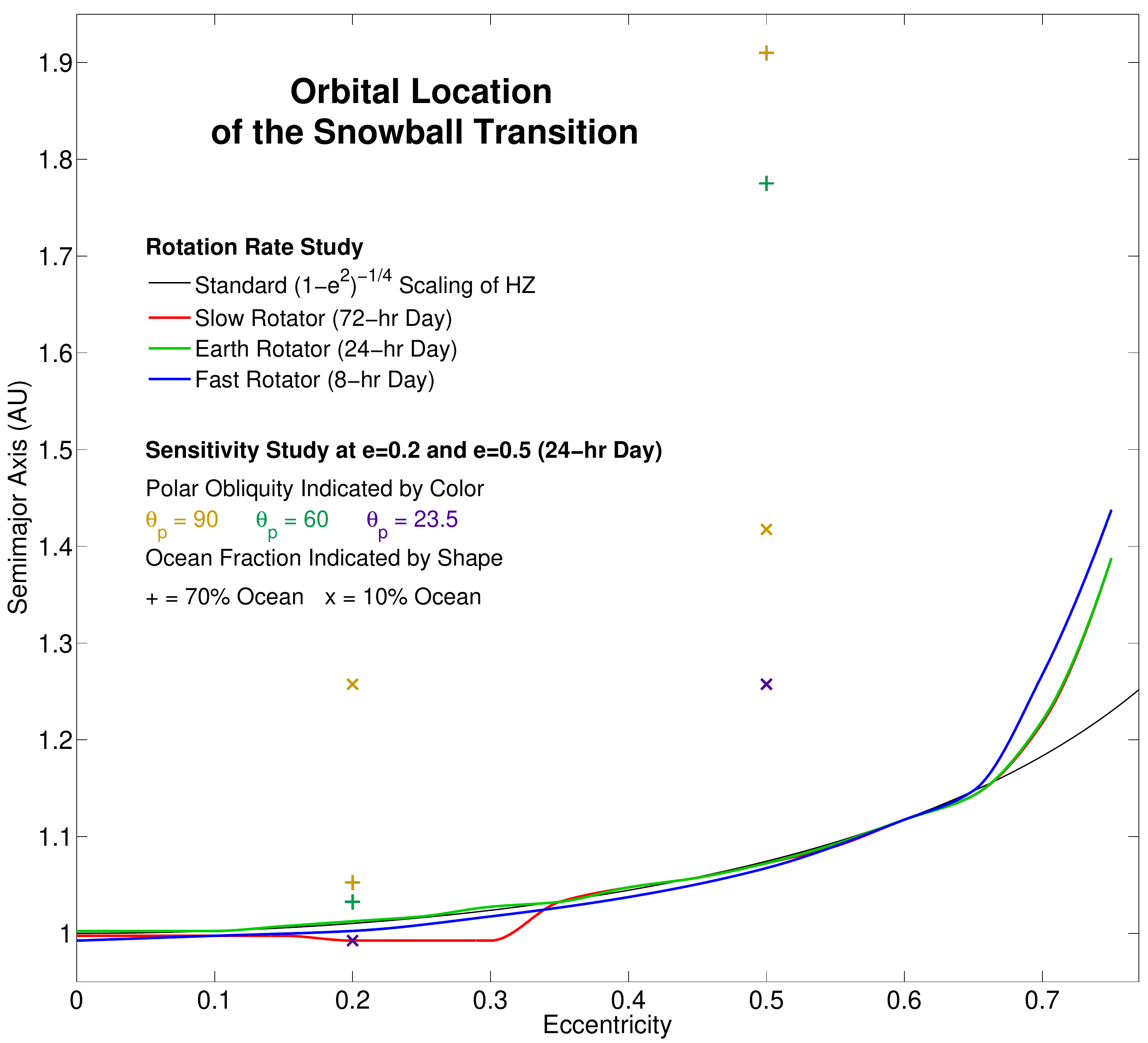}
\caption{Effect of rotation rate on the position of the snowball
  transition. Each line displays the approximate location of the
  snowball transition for planets with 72-hr days (red), 24-hr days
  (green), and 8-hr days (blue). For comparison, the symbols mark the
  position of the snowball transition for several of the model planets
  with 24-hr days used in the sensitivity study discussed in Section
  \ref{ssec:sense}. The orbital and planetary parameters for those
  planets are indicated above; for all other model planets
  ${\theta_p=23.5^\circ}$, ${\theta_a=0^\circ}$, and ${f_o=0.7}$. The
  black line shows where the (fixed-atmosphere) outer boundary of the
  habitable zone would be if it followed the $(1-e^2)^{-1/4}$ scaling
  that is sometimes assumed.  This scaling has been used by previous
  studies to estimate the position of the habitable zone, and matches
  well with our modeled habitable zones for eccentricities
  ${\lesssim0.65}$. Planets in more eccentric orbits appear to be
  habitable at greater semimajor axes than the simple scaling
  relationship would predict. For instance, the planet with
  ${\theta_p=90^\circ}$ and ${f_o=0.7}$ is habitable out to 1.90~AU.
  Rotation rate has little effect on the position of the snowball
  transition for eccentricities between 0.35 and 0.65, but increasing
  rotation rate appears to extend the position of the habitable zone
  for highly eccentric planets.  In low eccentricity orbits, the
  relationship between rotation rate and maximum habitable semimajor
  axis is not monotonic.  The complex relationship may indicate that
  the semimajor axis of the snowball transition is maximized when the
  planet rotates slowly enough that sufficient heat is transported to
  the mid-latitudes to prevent the advance of global permanent ice
  coverage but quickly enough that a significant amount of heat is
  retained in the low-latitudes.}
\label{fig:outerlim}
\end{figure}

\begin{figure}
\centering
\includegraphics[height=0.35\textwidth]{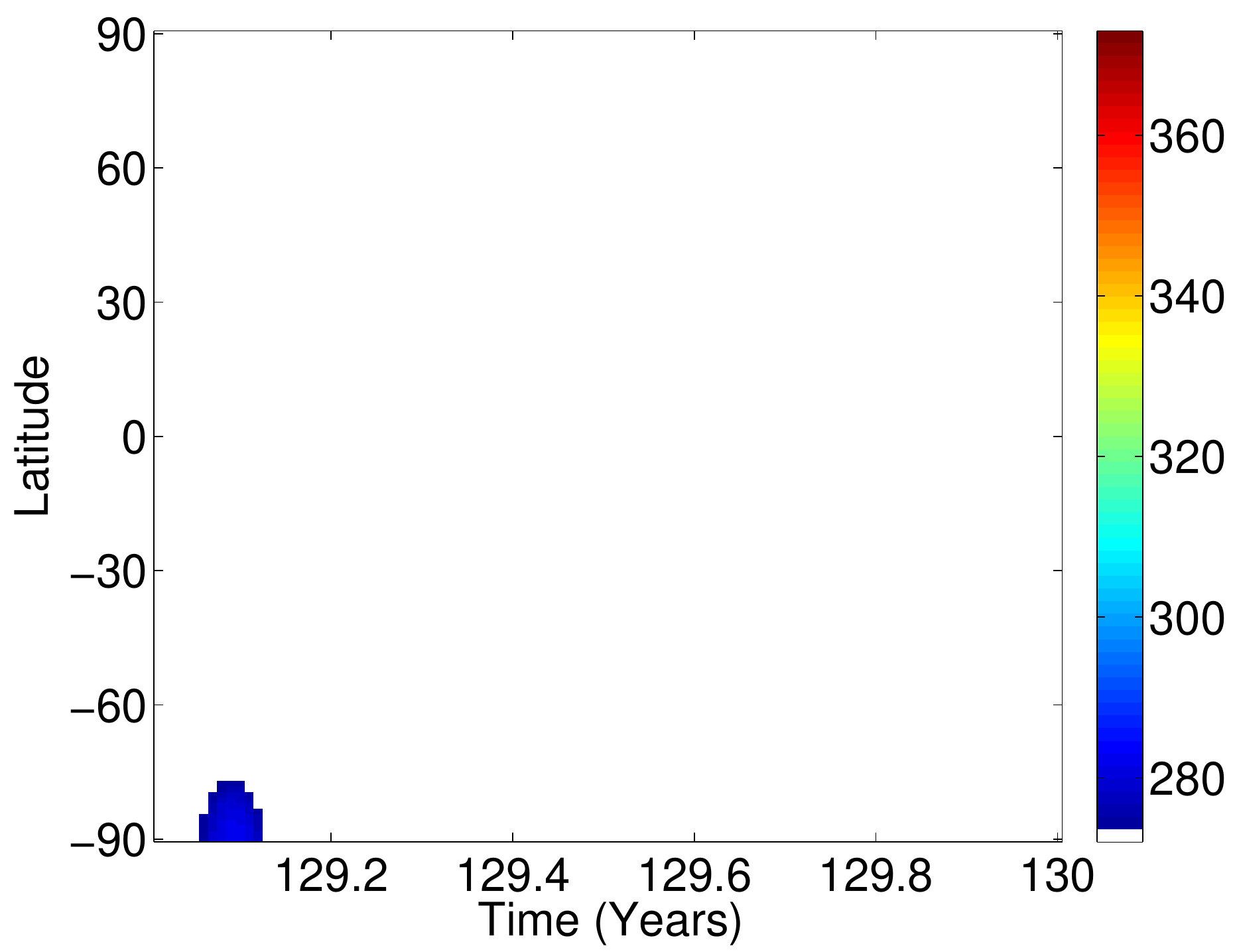}
\includegraphics [height=0.35\textwidth]{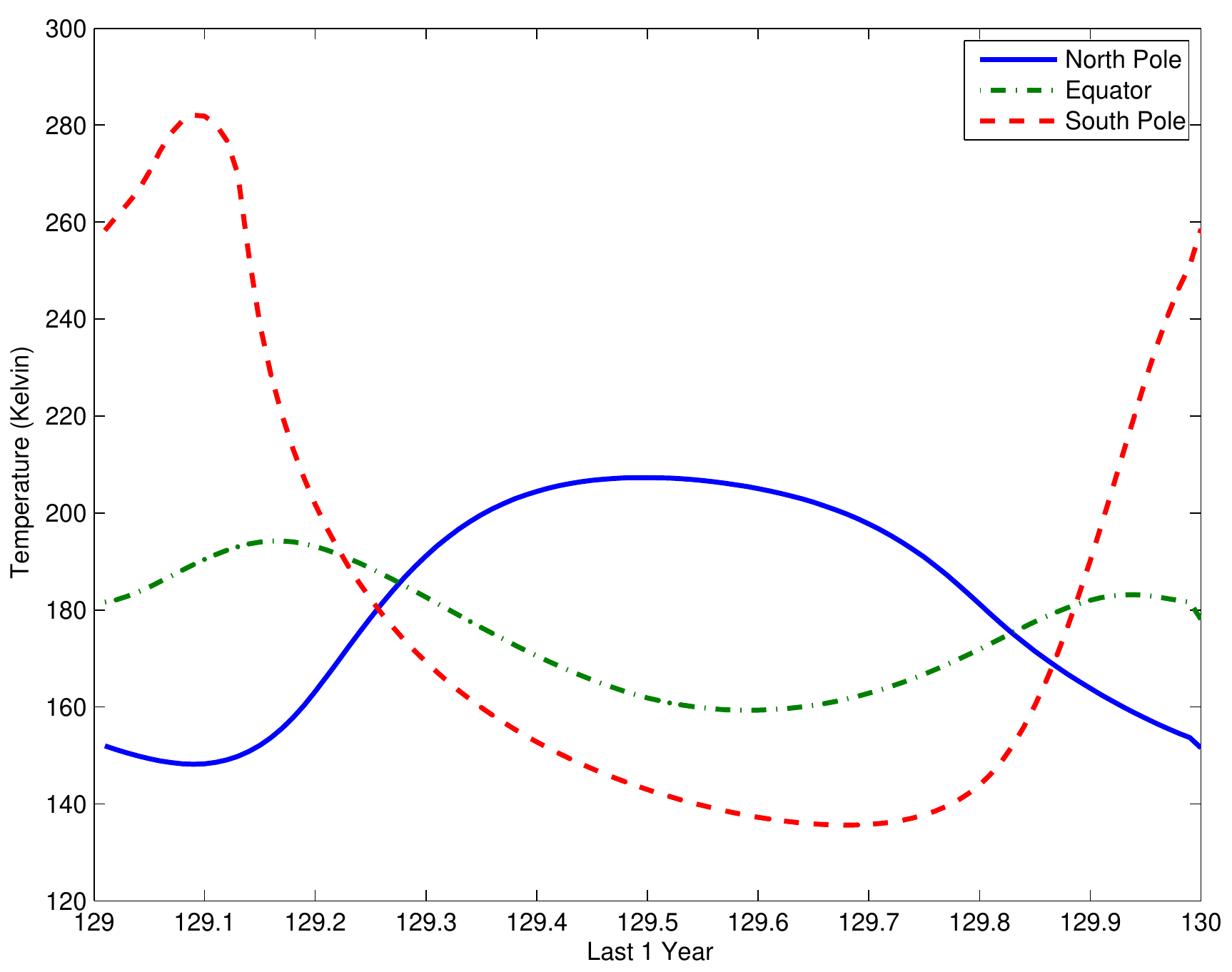}
\caption{Temperature profile of a pseudo-Earth with rotation rate
  $\Omega_p = \Omega_\earth$, azimuthal obliquity $\theta_a = 0$,
  polar obliquity $\theta_p = 90$, and ocean fraction = 10\% in an
  eccentric orbit with semimajor axis with $a = 1.255$~AU and
  eccentricity $e = 0.20$.  {\bf Left:} Habitability map of the planet
  with colors representing temperature as shown in the color bar.  The
  planet is completely frozen for most of the year, but the southern
  hemisphere warms to just above 273~K during southern summer at
  periastron.  {\bf Right:} Temperature variations during the course
  of the year for the latitude bands at the equator, north pole, and
  south pole.  Although the south pole is the only region of the
  planet that ever becomes habitable, it also experiences the most
  extreme temperature variations of any region on the planet.}
\label{fig:ee20verydryob90}
\end{figure}

\end{document}